\documentclass[twocolumn,superscriptaddress,showpacs,preprintnumbers,amsmath,amssymb]{revtex4-1}

\usepackage{amsmath}
\usepackage{amssymb}
\usepackage{graphicx}
\usepackage{morefloats}
\usepackage[usenames,dvipsnames]{xcolor}
\usepackage{blkarray}
\usepackage{verbatim}
\usepackage{hyperref}

\hypersetup{colorlinks=true, citecolor=orange, urlcolor=cyan, linkcolor=blue}

\begin{document}

\title{\bf{Randomizing bipartite networks: the case of the World Trade Web}}
\author{Fabio Saracco}
\affiliation{Istituto dei Sistemi Complessi (ISC) - CNR, UoS Sapienza, Dipartimento di Fisica, Universit\'a ``Sapienza'' di Roma, P.le A. Moro 5, 00185 Roma (Italy)}
\author{Riccardo Di Clemente}
\affiliation{Istituto dei Sistemi Complessi (ISC) - CNR, UoS Sapienza, Dipartimento di Fisica, Universit\'a ``Sapienza'' di Roma, P.le A. Moro 5, 00185 Roma (Italy)}
\author{Andrea Gabrielli}
\affiliation{Istituto dei Sistemi Complessi (ISC) - CNR, UoS Sapienza, Dipartimento di Fisica, Universit\'a ``Sapienza'' di Roma, P.le A. Moro 5, 00185 Roma (Italy)}
\affiliation{IMT Institute for Advanced Studies, P.zza S. Ponziano 6, 55100 Lucca (Italy)}
\affiliation{INFN - Unit\'a Roma1, Dipartimento di Fisica, Universit\'a ``Sapienza'' di Roma, P.le A. Moro 5, 00185 Roma (Italy)}
\author{Tiziano Squartini}
\affiliation{Istituto dei Sistemi Complessi (ISC) - CNR, UoS Sapienza, Dipartimento di Fisica, Universit\'a ``Sapienza'' di Roma, P.le A. Moro 5, 00185 Roma (Italy)}
\date{\today}

\begin{abstract}
Within the last fifteen years, network theory has been successfully applied both to natural sciences and to socioeconomic disciplines. In particular, bipartite networks have been recognized to provide a particularly insightful representation of many systems, ranging from mutualistic networks in ecology to trade networks in economy, whence the need of a pattern detection-oriented analysis in order to identify statistically-significant structural properties. Such an analysis rests upon the definition of suitable null models, i.e. upon the choice of the portion of network structure to be preserved while randomizing everything else. However, quite surprisingly, little work has been done so far to define null models for real bipartite networks. The aim of the present work is to fill this gap, extending a recently-proposed method to randomize monopartite networks to bipartite networks. While the proposed formalism is perfectly general, we apply our method to the binary, undirected, bipartite representation of the World Trade Web, comparing the observed values of a number of structural quantities of interest with the expected ones, calculated via our randomization procedure. Interestingly, the behavior of the World Trade Web in this new representation is strongly different from the monopartite analogue, showing highly non-trivial patterns of self-organization.
\end{abstract}

\maketitle

\section*{Introduction}

In the last fifteen years network science has exploded, revealing a world composed by interconnected systems ubiquitously found both in natural sciences and in socioeconomic disciplines \cite{Barabasi2002,Newman2003,Caldarelli2007}. Since the very beginning of network science, many different network representations have been adopted in order to study the particular system at hand \cite{Boccaletti2006}. However, the class of networks represented by bipartite networks has been recognized to provide a particularly insightful representation of many different systems \cite{Latapy2004}: ecological networks \cite{Dormann2009}, trade networks \cite{HH2009,Tacchella2012,Tacchella2013}, citations and collaboration networks \cite{Cimini2014,Alva2006} represent only few examples.

One could thus expect a relevant amount of work aimed at identifying the statistically-relevant patterns observed in real bipartite networks, at least comparable to the mass of results obtained so far for monopartite networks \cite{Chung2002,Caldarelli2002,Newman2004,Serrano2006,Garlaschelli2008,Bianconi2009,Fronczak2014,mymethod,myreconstruction,mysampling}: however, quite surprisingly, little work has been done so far to implement null models on real bipartite networks. Generally speaking, null models are statistical models used to make inference on a real system on the basis of partial information. The latter usually corresponds to some observable property of interest as the number of trade partners of a country, its exports and imports, the total exposure of a bank, etc. In particular, null models for bipartite networks being \emph{real-data rooted} and showing the desirable features of \emph{general applicability} and \emph{analytical character} are currently missing. More in detail, the algorithms proposed so far show several limitations, ranging from being purely numerical (thus lacking the analytical character) \cite{Dormann2009,Dormann2008,Strona2014}, to assuming an \emph{a priori} functional form either for the distribution of the quantities of interest \cite{Dormann2009} or for the model parameters (thus not being real data-rooted) \cite{Kitsak2011} or, lastly, using approximate analytical models \cite{Dormann2014}. Moreover, almost all the aforementioned approaches are tailored on ecological networks, thus lacking the character of general applicability.

The lack of such models is, maybe, also due to the misconception that bipartite networks could be analysed by, firstly, projecting them on one of the layers and, secondly, analysing the projection with one of the models currently available for monopartite networks. As we will show in what follows, the monopartite and the bipartite representations enclose different kinds of information, irreducible to each other (in the most general case).

The aim of the present paper is to fill this gap, proposing a theoretical framework guaranteeing the three aforementioned properties. In order to do this, we extend a recently-proposed method to randomize monopartite networks \cite{mymethod} to bipartite networks. The method rests upon the sequential maximizations of Shannon entropy and the network likelihood function, a combination which has been proven to be rather effective both for detecting patterns and to reconstruct the structure of several real-world networks \cite{myreconstruction,myPRE1,myPRE2,mymotifs,bootstrap,andrea}. To the best of our knowledge, the only other paper proposing a method satisfying the three requirements above is \cite{Tummi2011}: we will comment on the differences with the one proposed here in the Discussion section.

While the proposed formalism is perfectly general, in this paper we apply our method to the binary, undirected, bipartite representation of the World Trade Web (hereafter WTW). We focused on this particular system precisely because of its popularity among network scientists, who have applied null models to all its possible representations \cite{myPRE1,myPRE2,Fronczak2012,Serrano2003,Fagiolo2010,Barigozzi2010}, with the exception of the bipartite one. As we will show in what follows, representing the WTW as a bipartite network allows to gain a substantially new insight into an already deeply explored system.

The rest of the paper is organized as follows: Data section is devoted to the description of the dataset used for the present analysis, Methods section reports the detailed description of our method and Results section illustrates the results which are discussed in Discussion section, where conclusions are also drawn.

\section*{Data}

The WTW can be represented in many different ways, depending on the level of information that we want to process. The most popular ones represent it via an adjacency matrix with nodes playing the role of world-countries and links indicating the presence of (any kind) of trade exchange between them. This framework has been recently extended to analyse the WTW as a multiplex, where trade exchanges corresponding to different commodities are distinguished \cite{Barigozzi2010,myintensive}.

Here we represent the WTW as a bipartite network, i.e. by considering the set of world-countries and the set of products as different entities and linking a given country to a given product if (and only if) the former exports the latter \emph{above} a certain threshold (the so-called RCA \cite{Tacchella2012,Tacchella2013}). Applying the latter rises the probability that the exported commodity is actually produced by the exporting country. In this representation, any two countries (as well as any two products) cannot be directly linked (i.e. links connecting nodes of the same set are not allowed): thus, any two nodes of the same set can be still thought as ``interacting'' but only indirectly, via a connection with the same nodes of the other family. This way of representing the WTW allows us to analyze the global economy from a different perspective, by making the \emph{productivity relations} between countries explicit (i.e. \emph{which country produces which product}).

The dataset we have considered for the present analysis is the NBER database, collecting data for the 38 years 1963-2000 \cite{webdata} and categorizing products according to the SITC revision 2 at four-digits level. Data have been further processed, building upon the data-mining procedure adopted in \cite{Feenstra2005}, to produce a dataset with 538 products across all years and a number of countries varying from 130 to 151.

\section*{Methods}

\begin{figure}[t!]
\begin{center}
\includegraphics[width=0.49\textwidth]{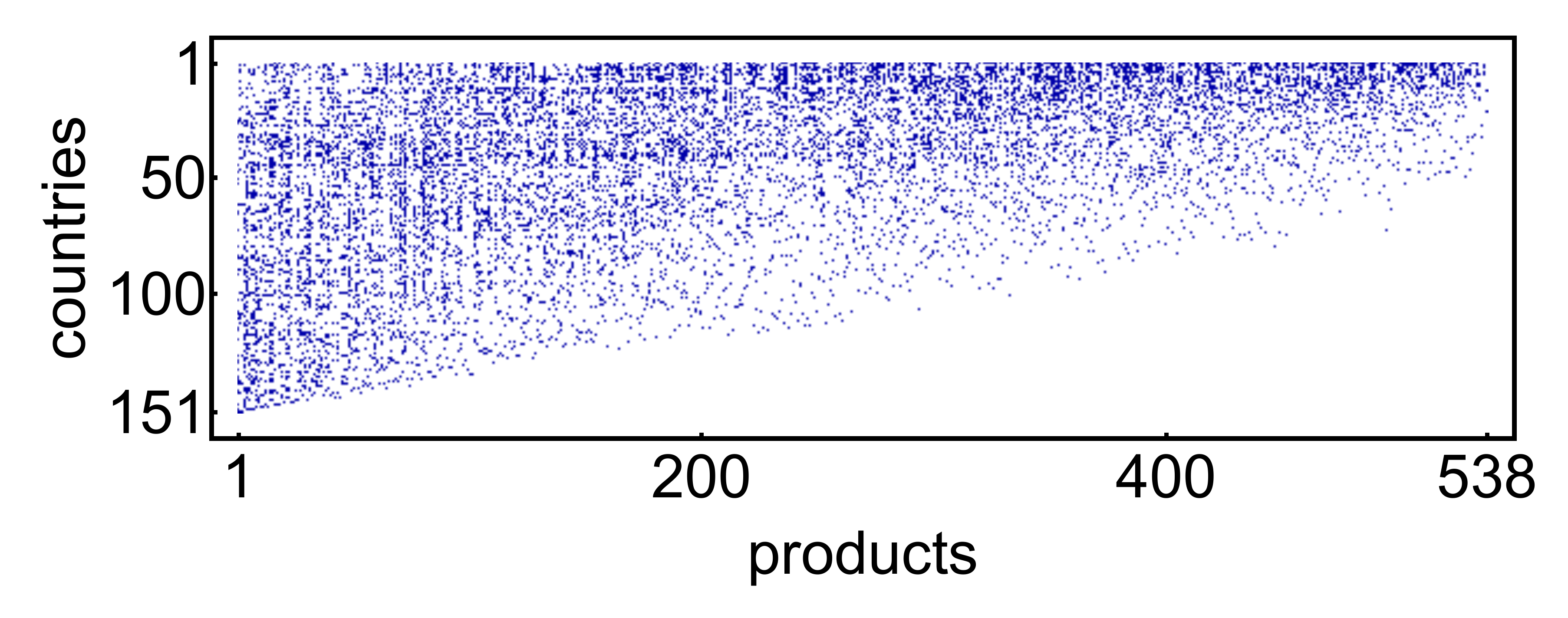}
\caption{The binary, undirected, bipartite representation of the World Trade Web in the year 2000 \cite{webdata}: countries are listed along the rows, products along the columns. Blue dots represent the ones, white dots represent the zeros. Rows and columns are reordered according to the algorithm introduced in \cite{Tacchella2012,Tacchella2013}.}
\label{one}
\end{center}
\end{figure}

The distinction between countries and products leads naturally to the definition of an biadjacency matrix, which will be indicated with $\mathbf{M}$. In the present paper we focus on the binary, undirected representation of the WTW: thus, the matrix entries will be either $m_{cp}=1$, indicating that country $c$ exports an amount of product $p$ above the RCA threshold, or $m_{cp}=0$, indicating that the production of $p$ by country $c$ is below the RCA threshold and, thus, has been ignored. As a consequence, each row represents the export basket of a given country, while each column represents the subset of producers of a given product. A pictorial representation of the WTW biadjacency matrix in the year 2000 is shown in fig. \ref{one}, with the blue dots representing the ones and the white dots the zeros.

If we indicate with $C$ the total number of countries and with $P$ the total number of products, the total number of elements of the biadjacency matrix (i.e. its volume) is $C\cdot P$, also representing the maximum observable number of connections. In fact, unlike the usual square representation, the problems arising from the presence of self-connections are not encountered here. Moreover, the presence of two different subsets (also known as \emph{layers}) induces a measure of ``rectangularity'' of our matrix $\mathbf{M}$ \cite{Dormann2009}, i.e. $R=\frac{|C-P|}{C+P}$, ranging in $R\in[0,1)$, with values closer to 1 indicating a large asymmetry between the number of countries and the number of products and values closer to 0 indicating equivalence between the two layers cardinality (notice that the information on the sign would be based on the arbitrary choice of the layers ordering).

The definitions of other topological quantities of interest easily follow from the usual ones, as the {\it number of links} (i.e. the total number of connections)

\begin{equation}
L(\mathbf{M})=\sum_{c=1}^C\sum_{p=1}^Pm_{cp},
\end{equation}

\noindent and the {\it connectance} $c(\mathbf{M})=\frac{L(\mathbf{M})}{C\cdot P}$, measuring the percentage of observed connections. Fundamental properties are represented by the number of node-specific connections, i.e. the {\it degree of countries}, also named \emph{diversification} \cite{HH2009,Tacchella2012,Tacchella2013}, measuring the number of products exported by each country

\begin{equation}
d_c(\mathbf{M})=\sum_{p=1}^Pm_{cp},
\label{cdeg}
\end{equation}

\noindent and the {\it degree of products}, also named \emph{ubiquity} \cite{HH2009,Tacchella2012,Tacchella2013}, measuring the number of countries exporting each product

\begin{equation}
u_p(\mathbf{M})=\sum_{c=1}^Cm_{cp}.
\label{pdeg}
\end{equation}

Definitions \ref{cdeg} and \ref{pdeg} induce the notions of {\it countries mean degree} and {\it products mean degree}

\begin{eqnarray}
\overline{d}(\mathbf{M})&=&\frac{\sum_{c=1}^C d_c(\mathbf{M})}{C}=\frac{L(\mathbf{M})}{C},\\
\overline{u}(\mathbf{M})&=&\frac{\sum_{p=1}^P u_p(\mathbf{M})}{P}=\frac{L(\mathbf{M})}{P}.
\label{eq.mean}
\end{eqnarray}

The last passage follows from noticing that $L(\mathbf{M})=\sum_{c=1}^Cd_c(\mathbf{M})=\sum_{p=1}^Pu_{p}(\mathbf{M})$.

In order to make the connections between nodes of the same family explicit, a bipartite network can be projected on its layers, thus recovering two traditional, monopartite representations. This operation can be straightforwardly implemented by considering the matrix products

\begin{equation}
\mathcal{C}=\mathbf{M}\cdot\mathbf{M^T},\:\mathcal{P}=\mathbf{M^T}\cdot\mathbf{M}
\end{equation}

\noindent where $\mathbf{M^T}$ is the transpose of the biadjacency matrix $\mathbf{M}$. While the dimensions of $\mathbf{M}$ are $C\times P$, the dimensions of its transpose are $P\times C$. This implies that $\mathcal{C}$ results in a $C\times C$ matrix whose generic element $\mathcal{C}_{cc'}$, with $c\neq c'$, counts the number of patterns of length two between countries $c$ and $c'$. The generic, diagonal element $\mathcal{C}_{cc}$ is precisely the degree of country $c$. Similarly, $\mathcal{P}$ results in a $P\times P$ matrix whose generic element $\mathcal{P}_{pp'}$, with $p\neq p'$, counts the number of patterns of length two between products $p$ and $p'$. As before, the generic, diagonal element $\mathcal{P}_{pp}$ is the degree of product $p$. Remarkably, the entries of matrices $\mathcal{C}$ and $\mathcal{P}$ have a clear macroeconomic interpreation: while $\mathcal{C}_{cc'}$ counts the number of products shared by countries $c$ and $c'$, $\mathcal{P}_{pp'}$ counts the number of countries exporting both products $p$ and $p'$.

Since nodes of the same layer cannot be directly linked, it is enough that a path of length two (i.e. the minimum allowed length) connects any two nodes of the same family to directly link them in the corresponding monopartite projection. Thus, by first applying the Heaviside step-function $\Theta[\dots]$ to matrices $\mathcal{C}$ and $\mathcal{P}$ element-wise (i.e. $\Theta[\mathcal{C}]=\{\Theta[\mathcal{C}_{cc'}]\}_{c,c'=1}^C$, where $\Theta[\mathcal{C}_{cc'}]$ can be 0 or 1, if $\mathcal{C}_{cc'}=0$ and $\mathcal{C}_{cc'}>0$ respectively - and similarly for $\mathcal{P}$) and then subtracting the diagonal elements, the binary, adjacency matrices describing the two monopartite projections are recovered, i.e.

\begin{equation}
\mathbf{C}=\Theta\left[\mathcal{C}\right]-\mathbf{I}_C,\:\mathbf{P}=\Theta\left[\mathcal{P}\right]-\mathbf{I}_P
\end{equation}

\noindent where $\mathbf{I}_C$ and $\mathbf{I}_P$ are the identity matrices having dimensions $C\times C$ and $P\times P$ respectively.

\subsection*{Topological measures for binary, undirected, bipartite networks}

Several quantities have already been proposed to analyse bipartite networks \cite{Dormann2009}. However, here we define different measures by extending some of the most used indicators in network theory, better capturing, in our opinion, the particular features of a given bipartite network structure.

\paragraph{Assortativity.} The traditional definition of assortativity is intended to quantify the degrees correlations, by distinguishing the assortative behavior (signalling positive degrees correlations) from the disassortative behavior (signalling negative degrees correlations). When dealing with bipartite networks, we can measure such correlations both with respect to countries and with respect to products, by respectively defining the {\it average nearest products ubiquity} (or ANPU)

\begin{equation}\label{eq:ANPDdef}
u^{nn}_c(\mathbf{M})=\frac{\sum_{p=1}^Pm_{cp}u_p}{d_c}
\end{equation}

\noindent and the {\it average nearest countries diversification} (or ANCD) as

\begin{equation}\label{eq:ANPDdef}
d^{nn}_p(\mathbf{M})=\frac{\sum_{c=1}^C m_{cp}d_c}{u_p}.
\end{equation}

As in the monopartite case, assortativity is quantified by respectively scattering the ANPU and ANCD values versus the degree sequences $\{d_c\}_{c=1}^C$ and $\{u_p\}_{p=1}^P$.

\paragraph{Complexity and fitness.} As recently pointed out \cite{Tacchella2012,Tacchella2013}, countries and products can be assigned two purely network-based quantities, known as \emph{fitness}, $F_c$ (to be assigned to countries), and \emph{complexity}, $Q_p$ (to be assigned to products), playing the role of non-monetary indicators of the economy development and providing a highly non-trivial way to rank the world-countries economic health (see also the Supplementary Information).

\paragraph{Motifs.} The usual clustering coefficient, measuring the hierarchical structure of a monopartite network, cannot be defined for bipartite networks: in fact, since no odd cycles of any length can be observed in bipartite networks (precisely because links within the same layer are forbidden) triangles cannot be observed as well; similarly, the usual triangular motifs cannot be defined \cite{Caldarelli2007,Milo2002}. 

\begin{figure}[t!]
\begin{center}
\includegraphics[width=0.49\textwidth]{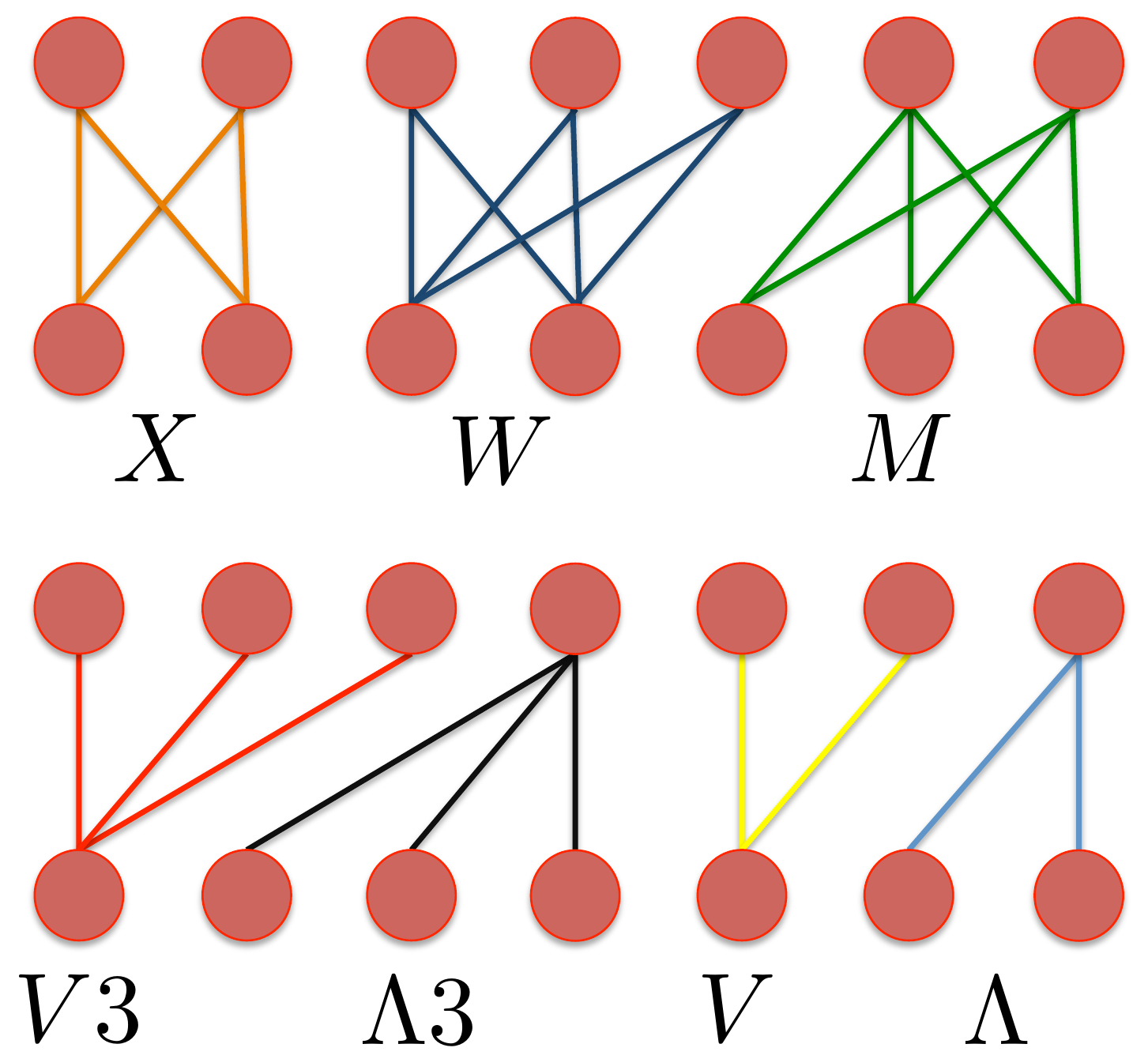}
\caption{Motifs for bipartite networks. Countries are reported in the upper layer, products in the bottom layer. The bottom panel shows motifs belonging to the V$n$ and $\Lambda n$ families, with $n=2, 3$.}
\label{motifs}
\end{center}
\end{figure}

However, higher-order correlations between nodes can still be captured by defining a completely new class of motifs. The first examples we provide are the V-\emph{motifs} and the \emph{$\Lambda$-motifs} (see fig. \ref{motifs}). The former count how many couples of countries export the same products, quantifying the productivities' similarity; the latter count how many couples of products are in the basket of the same producer, providing a measure of products correlation. Remembering that $\mathcal{C}_{cc'}$, with $c\neq c'$, counts the number of products exported by both $c$ and $c'$, the total number of V-motifs connecting any pair of countries is

\begin{eqnarray}
N_{V}(\mathbf{M})&=&\sum_{c=1}^C\sum_{c'=c+1}^C\mathcal{C}_{cc'}=\sum_{c=1}^C\sum_{c'=c+1}^C\sum_{p=1}^P m_{cp}m_{c'p}\nonumber\\&=&\sum_{p=1}^P \binom{u_p}{2}
\label{nvc}
\end{eqnarray}

\noindent and, remembering the analogous role of $\mathcal{P}_{pp'}$, the total number of $\Lambda$-motifs connecting any pair of products is

\begin{eqnarray}
N_{\Lambda}(\mathbf{M})&=&\sum_{p=1}^P\sum_{p'=p+1}^P\mathcal{P}_{pp'}=\sum_{p=1}^P\sum_{p'=p+1}^P\sum_{c=1}^C m_{cp}m_{cp'}\nonumber\\&=&\sum_{c=1}^C \binom{d_c}{2}.
\label{nvp}
\end{eqnarray}

The last passages follow from noticing that each V-motif ($\Lambda$-motif) is constituted by a pair of links having the same product (country) as a common vertex. The number of countries competing on the same product, as well as the number of products in the same basket, can be further risen, leading to the following generalizations (with V$2\equiv$V and $\Lambda 2\equiv \Lambda$):

\begin{equation}
N_{Vn}(\mathbf{M})=\sum_{p=1}^P \binom{u_p}{n},\:N_{\Lambda n}(\mathbf{M})=\sum_{c=1}^C \binom{d_c}{n};
\label{VL}
\end{equation}

\noindent fig. \ref{motifs} shows an example of V3-motifs and $\Lambda3$-motifs. From defintions (\ref{VL}) it follows that V1=$\Lambda1=L$.

Higher-order correlations can be captured by allowing for a higher number of connected nodes in the same layers (see X-\emph{motifs}, M-\emph{motifs} and W-\emph{motifs} in the Supplementary Information). Remarkably, all the defined kinds of motifs:

\begin{itemize}
\item can be compactly expressed in terms of products of biadjacency matrix entries; 
\item can be defined for specific subsets of countries and products, thus allowing for a finer analysis of the production dynamics. For example, a measure of correlation of countries $a$ and $b$ production is given by the motif $N_{V^{a,b}}=\mathcal{C}_{ab}=\sum_{p=1}^Pm_{ap}m_{bp}$;
\item may have an application also in the analysis of ecological networks, especially mutualistic networks (e.g. impollinators-flowers): in fact, measures of co-occurrence can be directly applied to ecosystems to quantify the species' competitiveness for the available resources.
\end{itemize}

In what follows we will focus on the V$n$ and $\Lambda n$ families (a more detailed discussion about all motifs is provided in the Supplementary Information). 

\paragraph{Assortativity coefficient.} Beside our definitons, we have also considered the assortativity measure proposed in \cite{Newman2002} and called $r$. The latter ranges in the domain $r\in[-1,1]$, with $r=1$ indicating the tendency of links to connect nodes with similar degrees and $r=-1$ indicating the tendency of links to connect nodes with different degrees.

\paragraph{Nestedness.} On the basis of the two aforementioned measures $F_c$ and $Q_p$, one can reorder the matrix rows and columns (i.e. countries and products) by, respectively, decreasing the fitness along rows (from top to bottom) and increasing the complexity along columns (from left to right), thus obtaining the triangular structure shown in fig. \ref{rnest}. In order to quantify the shape of such a matrix, several measures have been recently proposed \cite{Almeida2008,Bastolla2009,Allesina2013,Munoz2013}, under the common name of \emph{nestedness}. Here we adopt the one proposed in \cite{Almeida2008} (called NODF - see also the Supplementary Information). Notice that the measure of nestedness adopted here doesn't depend on the rows and columns ordering criterion (in what follows we will adopt the one based on $F_c$ and $Q_p$ measures) \cite{Tacchella2012,Tacchella2013}.

\subsection*{Randomizing bipartite networks}

In order to implement suitable null models to detect the statistically-relevant patterns of real bipartite networks, the lines of the method proposed in \cite{mymethod} can be followed. In particular, an ensemble $\mathcal{G}$ of binary, undirected, bipartite networks must be considered, in order to maximize Shannon entropy

\begin{equation}
S=-\sum_{\mathbf{M}\in\mathcal{G}}P(\mathbf{M})\ln P(\mathbf{M})
\end{equation}

\noindent under a given set of constraints $\vec{C}(\mathbf{M})$ \cite{Garlaschelli2008,mymethod}. Notice that the probability coefficient $P(\mathbf{M})$ is assigned to every adjacency matrices in the esemble and the constraints are defined in terms of the entries of $\mathbf{M}$. The result is the well-known exponential distribution:

\begin{equation}
P(\mathbf{M}|\vec{\theta})=\frac{e^{-H(\mathbf{M},\:\vec{\theta})}}{Z(\vec{\theta})}
\label{probcoef}
\end{equation}

\noindent with the hamiltonian $H(\mathbf{M},\:\vec{\theta})=\vec{\theta}\cdot\vec{C}(\mathbf{M})$ compactly expressing the imposed set of constraints, $\vec{\theta}$ being the vector of Lagrange multipliers associated to the vector of constraints and $Z(\vec{\theta})=\sum_{\mathbf{M}\in\mathcal{G}}e^{-H(\mathbf{M},\:\vec{\theta})}$ being the normalization. 

In the monopartite case, one of the most insightful null models has been proven to be the so-called Configuration Model (CM) \cite{Chung2002,Newman2004}. Let us now implement the bipartite extension of the CM (BiCM, in what follows), by constraining the degree sequence of the binary, undirected, bipartite WTW and analyzing the system beyond the information contained into it. Since now we have two different layers of nodes, the hamiltonian reads

\begin{equation}
H(\mathbf{M},\:\vec{\theta})=\vec{\alpha}\cdot\vec{d}(\mathbf{M})+\vec{\beta}\cdot\vec{u}(\mathbf{M}).
\end{equation}

Now we can calculate the probability coefficient (\ref{probcoef}), associating a probability to each network in the ensemble on the basis of the specific degree sequences $\vec{d}(\mathbf{M})$ and $\vec{u}(\mathbf{M})$:

\begin{eqnarray}
P(\mathbf{M}|\vec{\theta})&=&\frac{e^{-\vec{\alpha}\cdot\vec{d}(\mathbf{M})-\vec{\beta}\cdot\vec{u}(\mathbf{M})}}{\sum_{\mathbf{M}}e^{-\vec{\alpha}\cdot\vec{d}(\mathbf{M})-\vec{\beta}\cdot\vec{u}(\mathbf{M})}}\nonumber\\&=&\prod_{c,p}p_{cp}^{m_{cp}}(1-p_{cp})^{1-m_{cp}}
\label{probcoef2}
\end{eqnarray}

\noindent the notation $\Pi_{c,p}$ being equivalent to $\Pi_{c=1}^C\Pi_{p=1}^P$ (see the Supplementary Information for the detailed calculations). The coefficient $p_{cp}=\frac{x_cy_p}{1+x_cy_p}$, with $e^{-\alpha_c}= x_c$ and $e^{-\beta_p}= y_p$, is the ensemble probability of having a link between country $c$ and product $p$, as $\langle m_{cp}\rangle=\sum_{\mathbf{M}\in\mathcal{G}}m_{cp}(\mathbf{M})P(\mathbf{M}|\vec{\theta})=p_{cp}\equiv\frac{x_cy_p}{1+x_cy_p}$. 

Our null model provides the analytical expression of a network probability as a product over all the accessible $C\times P$ pairs of nodes. In other words, the BiCM interprets the links as independent random variables, thus defining a grandcanonical probability measure where links correlations are discarded. Notice also that no probability coefficients controlling for the presence of links between nodes in the same layer appear in the expression (\ref{probcoef2}). This is a consequence of having considered an \emph{ensemble of bipartite networks} as the support of our probability distribution: in so doing, the forbidden intra-layer links are automatically excluded by the choice of the allowable configurations volume. 

The probability distribution in (\ref{probcoef2}) depends on $C+P$ unknown parameters (i.e. the Lagrange multipliers), also called {\it hidden variables} \cite{Caldarelli2002,Kitsak2011}. The recipe provided by statistical mechanics to estimate the hidden variables is summed up by the equations

\begin{equation}
-\frac{\partial\ln Z}{\partial \alpha_c}=\langle d_c\rangle,\:\forall\:c;\:\:-\frac{\partial\ln Z}{\partial \beta_p}=\langle u_p\rangle,\:\forall\:p.
\end{equation}

However, no indication about the numerical value to be assigned to the ensemble average of constraints is provided. Thus, in order to estimate the hidden variables from data, let us first note that $P(\mathbf{M}|\vec{\theta})$ can be rewritten solely in terms of the observed constraints value, i.e. $P(\mathbf{M}|\vec{\theta})=\prod_{c}x_c^{d_c(\mathbf{M})}\prod_{p}y_p^{u_p(\mathbf{M})}\prod_{c,p}\left(1+x_cy_p\right)^{-1}$ \cite{mymethod}. Then, let us consider the log-likelihood function $\mathcal{L}(\vec{x},\:\vec{y})=\ln P(\mathbf{M}|\vec{x},\:\vec{y})$:

\begin{eqnarray}
\mathcal{L}(\vec{x},\:\vec{y})&=&\sum_{c=1}^{C}d_c(\mathbf{M})\ln x_c+\sum_{p=1}^{P}u_p(\mathbf{M})\ln y_p+\nonumber\\&-&\sum_{c=1}^C\sum_{p=1}^P\ln(1+x_cy_p).
\end{eqnarray}

The recipe provided by statistics to estimate the unknown parameters of a given probability distribution prescribes to maximize $\mathcal{L}$ \cite{mymethod}. This means solving the system $\vec{\nabla}\mathcal{L}(\vec{x},\:\vec{y})=\vec{0}$ of $C+P$ equations in $C+P$ unknowns \cite{mymethod}:

\begin{equation}
\left\{ \begin{array}{ll}
d_c(\mathbf{M})&=\sum_{p=1}^P\frac{x_cy_p}{1+x_cy_p},\:c=1\dots C,\\
u_p(\mathbf{M})&=\sum_{c=1}^C\frac{x_cy_p}{1+x_cy_p},\:p=1\dots P.
\end{array}
\right.\\
\label{sys}
\end{equation}

In what follows the vector of solutions satisfying the system (\ref{sys}), for given $\vec{d}(\mathbf{M})$ and $\vec{u}(\mathbf{M})$ as degree mean values, will be indicated as $(\vec{x}^*,\:\vec{y}^*)$. Notice that the coefficients appearing at the second member of the system equations have the same functional form both for countries and products. This is a consequence of assigning only one Lagrange multiplier to each node but in such a way to distinguish the nodes in the first layer from the nodes in the second layer.

\subsection*{Expected topological measures for binary, undirected, bipartite networks}

In the previous subsections several quantities of interest to be measured on binary, undirected, bipartite networks have been listed. In this subsection we will show how our method can be implemented to calculate their expected value (to be compared with the observed one) and the relative errors (to quantify the discrepancies) in order to assess up to what level our null model is able to explain the higher-order structure of the network.

Our method allows us to proceed in a two-fold way. The first one is analytical. Using the link-specific probability coefficients $p_{cp}$ and the passages sketched in \cite{mymethod}, we are able to analytically calculate both the expected value and the standard deviation of the (analytically-definable) quantities of the previous subsections. However, because of the impossibility to perform analytical evaluation of the average for some key quantities, we have adopted a different strategy: we have sampled the grancanonical ensemble of binary, undirected, bipartite networks induced by the BiCM according to the probability coefficients $P(\mathbf{M}|\vec{x}^*,\:\vec{y}^*)$, measured the aforementioned properties on our sample $\tilde{\mathcal{G}}$ and calculated the statistical moments, as average and standard deviation, of the generic quantity $X$ as

\begin{eqnarray}
\langle X\rangle&\simeq&\tilde{X}=\sum_{\mathbf{M}\in\tilde{\mathcal{G}}}X(\mathbf{M})\tilde{P}(\mathbf{M}),\\
\sigma_X&\simeq&\sigma_{\tilde{X}}=\sum_{\mathbf{M}\in\tilde{\mathcal{G}}}(X(\mathbf{M})-\tilde{X})^2\tilde{P}(\mathbf{M})\nonumber\\
\end{eqnarray}

\noindent i.e. as sampling moments according to the sampling frequencies $\tilde{P}(\mathbf{M})=\frac{N_\mathbf{M}}{|\tilde{\mathcal{G}}|}$ ($N_\mathbf{m}$ being the number of networks in the ensemble having biadjacency matrix equal to $\mathbf{m}$). Since our method is unbiased \cite{mymethod,mysampling}, numerically sampling $\mathcal{G}$ provides a faithful representation of the whole ensemble. We have also calculated the probability distribution (induced by $\tilde{P}(\mathbf{M})$) of some of the properties of interest, in order to quantify the statistical significance of their observed value (via the $z$-score, for example). 

Nevertheless, the analytical expressions of the expected value and standard deviation of the quantities explicitly defined in the previous subsections has been derived in the Supplementary Information.

\section*{Results}

Let us first show our results on the temporal snapshot of the WTW corresponding to the year 2000. The number of nodes is $C_{2000}=151$ and $P_{2000}=538$, causing the $R$ index to be $R\simeq 0.56$ (see section Methods). The high asymmetry of our network is also pointed out by the different mean degrees, $\overline{d}\simeq 70$ and $\overline{u}\simeq 20$, indicating that countries are, on average, almost three times more connected than products. However, the connectance is $c_{2000}\simeq0.13$: thus, our bipartite WTW is much sparser than its monopartite counterpart \cite{myPRE1}. Notice that our null model, constraining (on average) the degree sequence, exactly reproduces any network's connectance by definition, spanning the domain of applicability of both the sparse and the dense network reconstruction algorithms.

\begin{figure*}[t!]
\begin{center}
\includegraphics[width=1\textwidth]{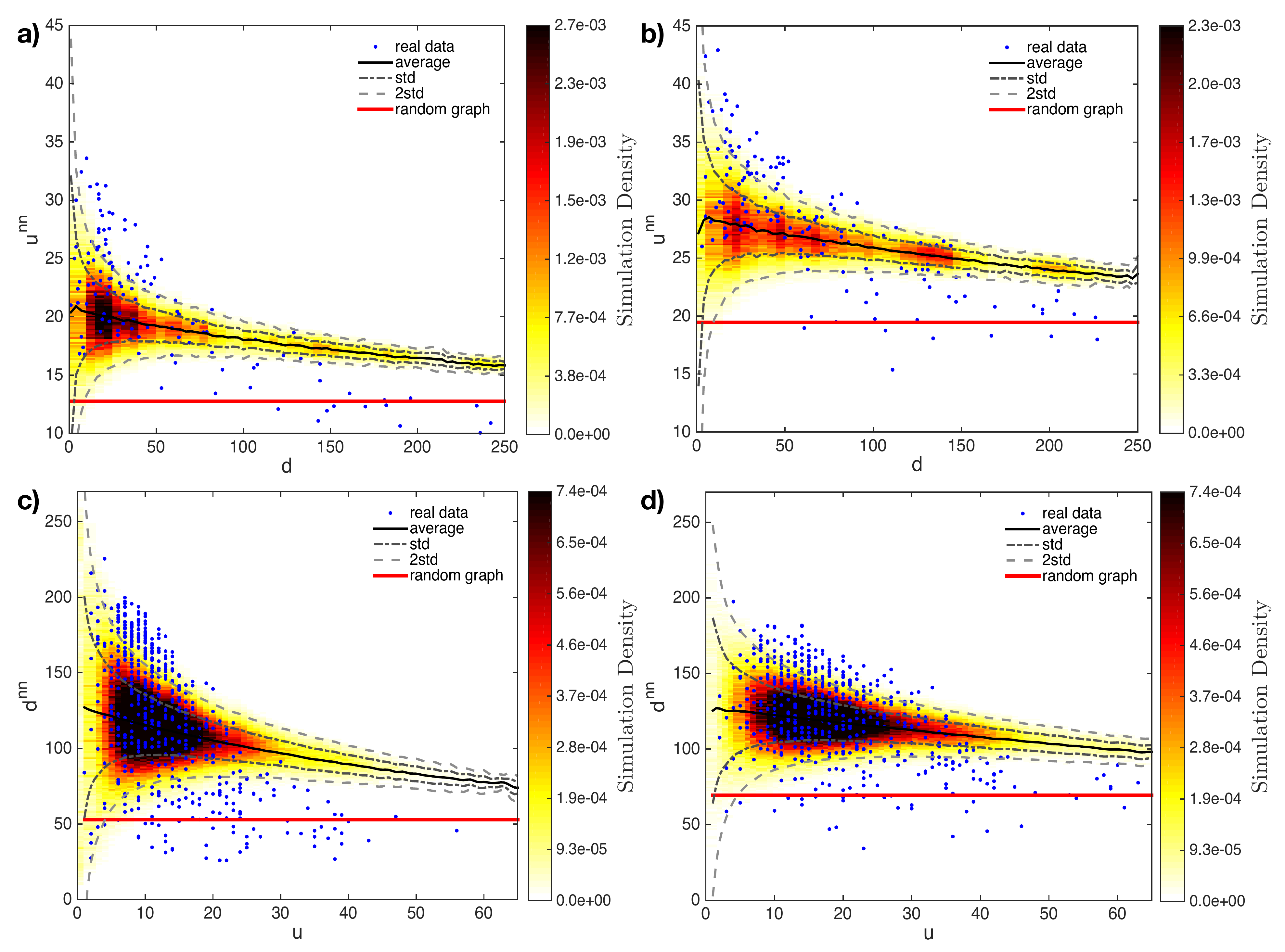}
\caption{Application of our method to the binary, undirected, bipartite World Trade Web in the year 1963 (left column) and 2000 (right column). Panels report $u^{nn}_c$ VS $d_c$ (a, b) and $d^{nn}_p$ VS $u_p$ (c, d). Observed points are in blue; the black, solid curves are CM-induced ensemble averages; the red, solid lines are RG-induced ensemble averages; the gray, dashed curves indicate the $\pm 1$ standard deviation region; the gray, dash-dotted curves indicate the $\pm 2$ standard deviations region. Colored areas represent the ensemble density of expected points (sampling 5000 matrices). Although the BiCM captures the disassortative trend of the WTW, its striking similarity with the BiRG predictions proves that the explanatory power of the degree sequence is far more limited in the bipartite representation than in the monopartite one \cite{myPRE1}.}
\label{knn2000}
\end{center}
\end{figure*}

\subsection*{Assortativity} 

Fig. \ref{knn2000} shows the comparison between observed and expected values of our coefficients of assortativity. Having plotted $u^{nn}_c$ VS $d_c$ and $d^{nn}_p$ VS $u_p$, we firstly observe that the bipartite WTW shows a \emph{disassortative} behavior, signalled by a globally decreasing trend of our measures. More detailedly, two distinct behaviors seem to characterize $u^{nn}_c$ as a function of $d_c$: while countries with \emph{low diversification} are preferentially linked to products with \emph{high ubiquity} (left side of panels \ref{knn2000}a and \ref{knn2000}b), countries with \emph{high diversification} are linked to \emph{almost all products} (right side of panels \ref{knn2000}a and \ref{knn2000}b). This is also reflected in the triangular structure of the matrix (see fig. \ref{one}). For products, this distinction is less sharp (panels \ref{knn2000}c and \ref{knn2000}d): in fact, while \emph{high-ubiquity} products are linked to \emph{almost all countries}, \emph{low-ubiquity} products can be found connected to \emph{both high-} and \emph{low-diversification countries}.

\begin{figure*}[t!]
\begin{center}
\includegraphics[width=0.99\textwidth]{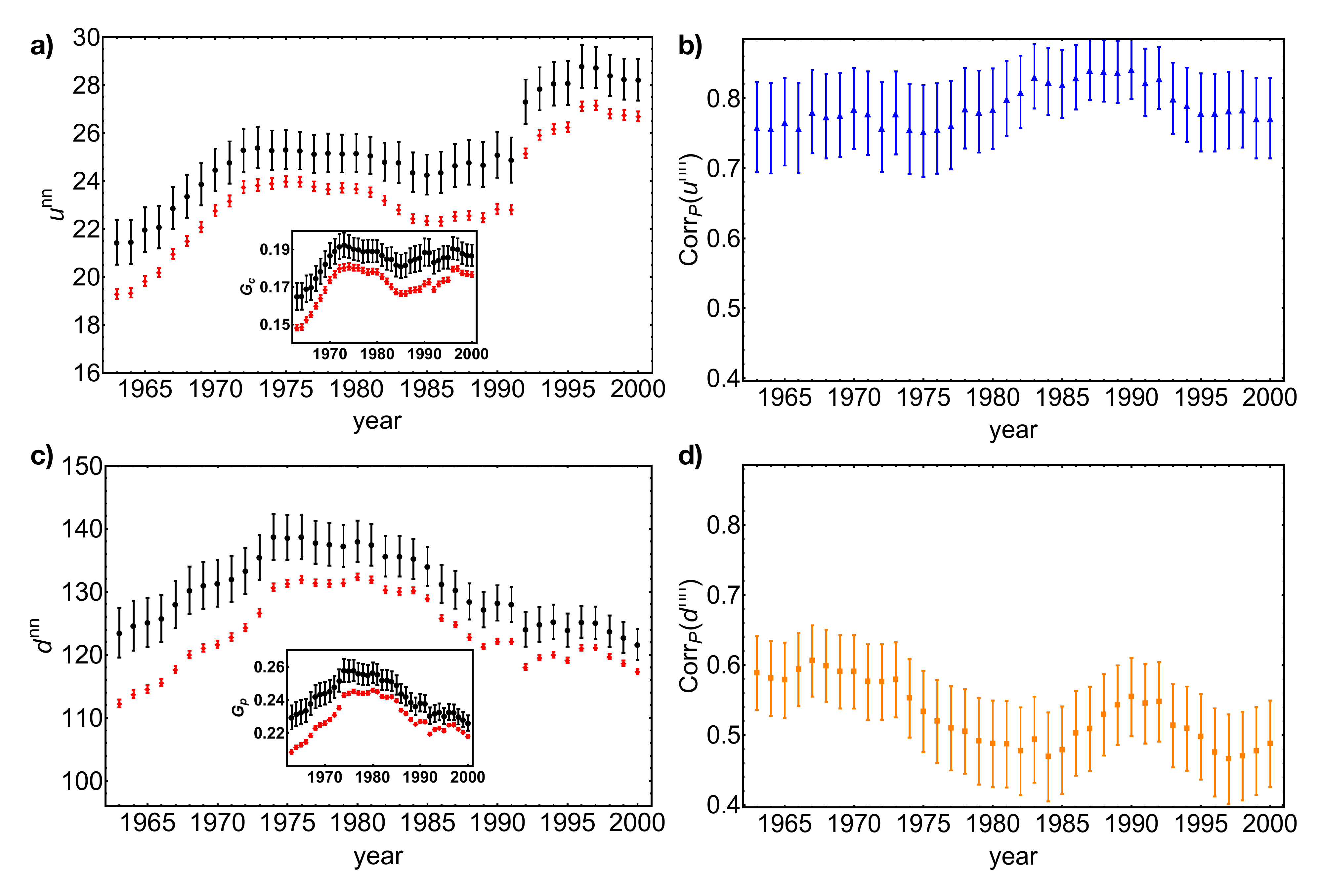}
\caption{Temporal evolution of the arithmetic mean of the observed $\{u^{nn}_c\}_{c=1}^C$ ($\textcolor{black}{\bullet}$) and expected $\{\langle u^{nn}_c\rangle\}_{c=1}^C$ ($\textcolor{red}{\bullet}$), together with the 95\% CI (panel a); temporal evolution of the arithmetic mean of the observed $\{d^{nn}_p\}_{p=1}^P$ ($\textcolor{black}{\bullet}$) and expected $\{\langle d^{nn}_p\rangle\}_{p=1}^P$ ($\textcolor{red}{\bullet}$), together with the 95\% CI (panel c); temporal evolution of the Pearson correlation coefficient between $\{u^{nn}_c\}_{c=1}^C$ and $\{\langle u^{nn}_c\rangle\}_{c=1}^C$ ($\textcolor{blue}{\blacktriangle}$) together with the 95\% CI (panel b) and between $\{d^{nn}_p\}_{p=1}^P$ and $\{\langle d^{nn}_p\rangle\}_{p=1}^P$ ($\textcolor{orange}{\blacksquare}$) together with the 95\% CI (panel d). The evolution of expected points closely follows the evolution of the observed ones, pointing out that the BiCM correctly describes the temporal trend of the assortativity indices.}
\label{knnevot}
\end{center}
\end{figure*}

As can be seen from fig. \ref{knn2000}, the BiCM captures the disassortative behavior of both $u^{nn}_c$ and $d^{nn}_p$; however, only part of the observed points lies within the $\pm 2$ standard deviations region. This means that the mechanism shaping the disassortative behavior of the WTW is not completely explained by our null model, signalling a non-trivial origin of the WTW degree correlations. What is strikingly surprising is the prediction based on the Random Graph model (BiRG): the corresponding trend is closer to the BiCM prediction than in the monopartite representation of the WTW \cite{myPRE1}. Moreover, since disassortativity is more pronounced in real data, our results indicate that the BiCM performs better than BiRG for small values of $d_c$ and $u_p$, while the BiRG correctly capture their flat behavior at large $d_c$ and $u_p$ (i.e. for competitive countries and ubiquitous products, for which $\langle d^{nn}_p\rangle_{BiRG}\simeq L/C$, $\langle u^{nn}_c\rangle_{BiRG}\simeq L/P$). This seems to indicate that the explanatory power of the degree sequence is far more limited in the bipartite representation than in the monopartite one and that additional information is required to improve the agreement between observations and predictions (even at the simplest level of binary, undirected networks).

Fig. \ref{knnevot} extends our assortativity analysis to the entire dataset. In order to condensate the information of 38 scatter plots, we have computed the barycenter and sparseness of both the observed and expected clouds of points. In particular, we have calculated the arithmetic mean of both the observed values $\{u^{nn}_c\}_{c=1}^C$

\begin{eqnarray}
\overline{u^{nn}}&=&\frac{\sum_{c=1}^C u_c^{nn}}{C}=\frac{1}{C}\sum_{c=1}^C \sum_{c'=1}^C\frac{\mathcal{C}_{cc'}}{d_c}
\end{eqnarray}

\noindent  and $\{d^{nn}_p\}_{p=1}^P$, the expected values $\{\langle u^{nn}_c\rangle\}_{c=1}^C$ and $\{\langle d^{nn}_p\rangle\}_{p=1}^P$ and the corresponding confidence intervals (CI) at 95\% level. As for the motifs, also $\overline{u^{nn}}$ and $\overline{d^{nn}}$ can be interpreted in macroeconomic terms. In fact, $\sum_{c'=1}^C\mathcal{C}_{cc'}/d_c$ measures the country-specific number of competitions, thus quantifying the (average) presence of a country on the global market. Further averaging over all countries provides a measure of the integration of world-countries production. 

\begin{figure*}[t!]
\begin{center}
\includegraphics[width=1\textwidth]{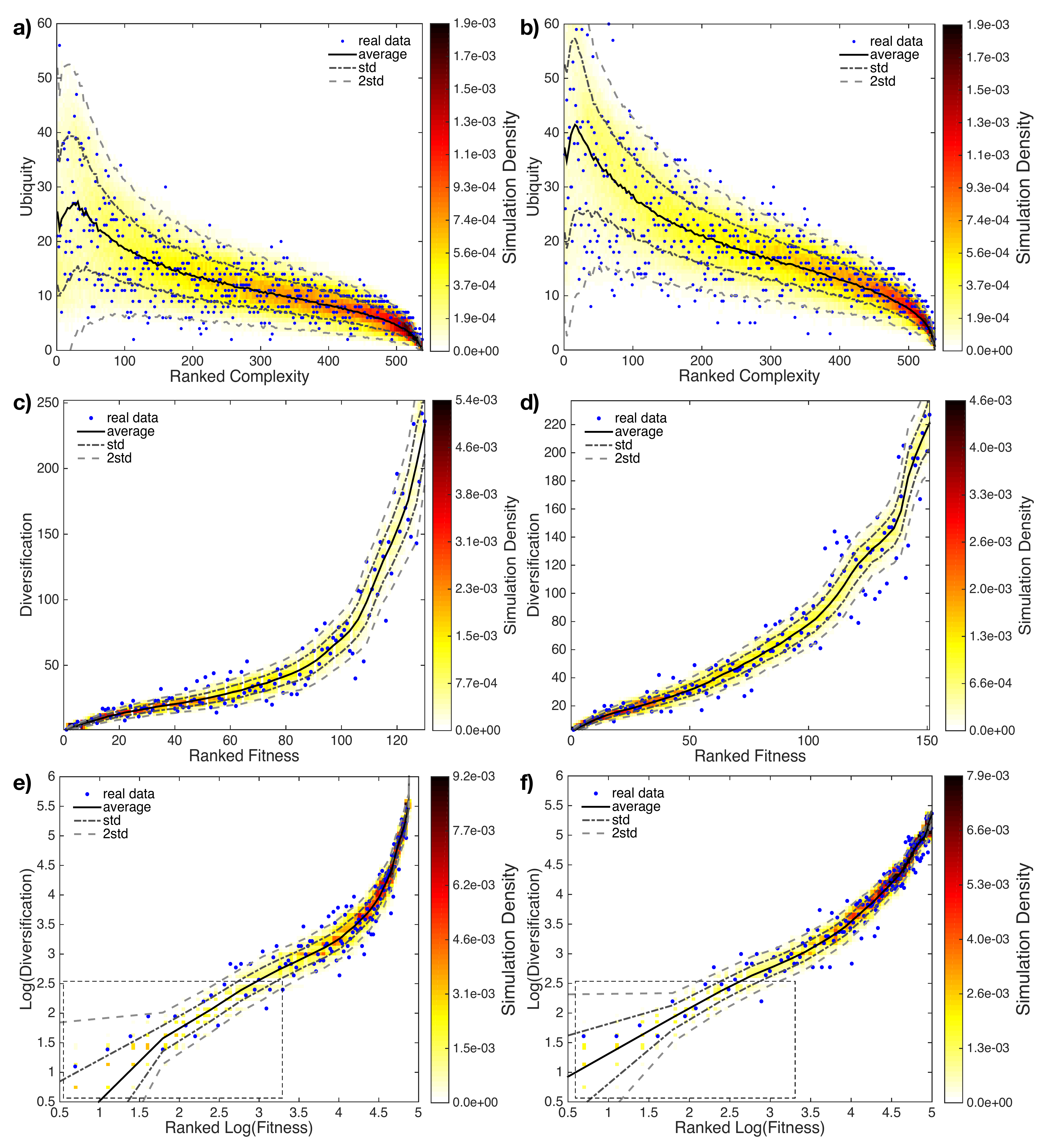}
\caption{Application of our method to the binary, undirected, bipartite World Trade Web in the year 1963 (left column) and 2000 (right column). Panels report $u_p$ VS $Q_p$ (a, b) and $d_c$ VS $F_c$ (c, d). Observed points are in blue; the black solid curves are BiCM-induced ensemble averages; the gray dashed curves indicate the $\pm 1$ standard deviation region; the gray dash-dotted curves indicate the $\pm 2$ standard deviations region. Colored areas represent the ensemble density of expected points (sampling 5000 matrices). Our null model seems to satisfactorily capture both trends. Panels (e, f) show the so-called ``poverty trap'', i.e. the group of countries with lowest fitness \cite{Tacchella2012,Tacchella2013}. Notice how all such countries lie within the $\pm 2$ standard deviation region (or immediately outside).}
\label{qf2000}
\end{center}
\end{figure*}

What emerges is that the evolution of expected points closely follows the evolution of the observed ones, pointing out that the BiCM correctly describes the temporal trend of the assortativity measures. Notice that, even if observed points are systematically more concentrated on higher levels (as shown in panels \ref{knnevot}a and \ref{knnevot}c), the confidence intervals are still close enough to let us interpret the BiCM predictions as correct. Moreover, the constancy of the amplitude of the confidence intervals for both observed and expected ANPU values indicates that the corresponding clouds of points maintain the same sparseness across our 38 years dataset; on the other hand, the amplitude of the observed ANCD confidence intervals slightly reduces, indicating a shrinkage of the corresponding cloud of points (compare panels \ref{knn2000}b and \ref{knn2000}d).

The temporal trends of $\overline{u^{nn}}$ and $\overline{d^{nn}}$ show interesting differences. In fact, while $\overline{u^{nn}}$ keeps increasing across the whole dataset, $\overline{d^{nn}}$ does not (and from 1975 starts decreasing). Since the countries mean degree keeps rising as well ($\overline{d}_{1963}\simeq48$ and $\overline{d}_{2000}\simeq70$), the increasing trend is probably due to the birth of new links, indicating that while existing countries have enlarged their production, new-born countries have started theirs. The results seem also to be compatible with the picture of several ``appealing'' products behaving as hubs and attracting links, including the ones of the new-born countries which in turn, having a low degree, reduce the value of the $d^{nn}_p$.

Since $\overline{u^{nn}}$ ranges in the interval $[0, C]$, the effect due to the varying number of countries can be washed away by further dividing it by $C$, $G_C=\overline{u^{nn}}/C$ (and thus normalizing it to the interval $[0,1]$). Remarkably, our index $G_C$ can be now interpreted a ``genuine'' measure of globalization, not affected by any spurious effect. Very interestingly, the temporal trend of $G_C$ after 1970 becomes now almost flat. This means that the WTW evolution does not actually affect the value of countries integration which organize in such a way to maintain the same value of $G_C$, irrespectively of the rising number of countries, their higher diversification, etc. This seems to confirm the stationary evolution of such network, recently pointed out \cite{mystationary}. A similar reasoning leads us to interpret $G_P=\overline{d^{nn}}/P$ as a measure of products homogeneity.

We have also calculated the Pearson correlation coefficient between the vectors $\{u^{nn}_c\}_{c=1}^C$ and $\{\langle u^{nn}_c\rangle\}_{c=1}^C$ (panel \ref{knnevot}b) and between the vectors $\{d^{nn}_p\}_{p=1}^P$ and $\{\langle d^{nn}_p\rangle\}_{p=1}^P$ (panel \ref{knnevot}d), in order to quantify the agreement on the ``shape'' of the clouds of points. The correlation of the latter is lower than the correlation of the former: this is due to the shape of the empirical cloud of ANCD which is less linear than the empirical ANPU, thus worsening the agreement with the corresponding expectations (which show an almost perfectly linear trend).

\subsection*{Complexity and fitness} 

Complexity and fitness can be obtained only numerically, as the result of the convergence of the algorithm proposed in \cite{Tacchella2012,Tacchella2013,Pugliese2014}. Panels \ref{qf2000}a and \ref{qf2000}b show the comparison between observed and expected complexity (plotted VS ubiquity) for the years 1963 and 2000; panels \ref{qf2000}c and \ref{qf2000}d show the comparison between observed and expected fitness (plotted VS diversification) for the same years. Our null model capture both trends with a larger accuracy than in the measure of assortativity: notice how the expected trend under the BiCM reproduces the ``beak'' of the observed complexity in real data and the vast majority of the observed cloud lies within the $\pm 2$ standard deviations region. 

Similarly, the expected trend of reconstructed fitness captures the different growth regimes of the observed fitness in the WTW data, showing few sparse points outside the same error region (clearly visible in the log-log plots of fig. \ref{qf2000}). The regime with lower slope (left side of panels \ref{qf2000}e and \ref{qf2000}f) represents the so-called ``poverty trap'' \cite{Tacchella2012,Tacchella2013}, i.e. the area populated by the group of countries with lowest fitness: notice how all such countries lie within the $\pm 2$ standard deviation region (or immediately outside). Similar considerations hold for all the remaining years, indicating a constant performance of our method across our 38-years dataset.

The average trends in fig. \ref{qf2000} are computed differently from those in fig. \ref{knn2000}: while the latter represent the node-specific, ensemble averages $\{\langle d_c^{nn}\rangle\}_{c=1}^C$ and $\{\langle u_p^{nn}\rangle\}_{p=1}^P$, the former represent averages taken over ranked nodes, ordered according to their complexity - panels a and b - and fitness - panels c and d. Generally speaking, ordering nodes on the basis of such procedure will produce a different ranking for different bipartite networks of the ensemble. Moreover, the ranking operation guarantees neither that the identity of ranked nodes remains the same (e.g. two different countries can be ranked first for two different networks), nor that the corresponding complexity and fitness maintain their value across our sample (i.e. the nodes ranked first will, in general, have different values of $F_c$ and $Q_p$): this in turn implies that each ranked node degree may change as well (i.e. the nodes ranked first for different networks will, in general, have different degrees). From these considerations, the need of quantifying 1) the variation of any country diversification as a function of its fitness and 2) the variation of any product ubiquity as a function of its complexity follows. This is in line with the spirit of the research in \cite{Tacchella2012,Tacchella2013}: trying to establish a biunivocal relation both between ubiquity and complexity and between fitness and diversification, in order to unambiguously rank countries and products. This kind of analysis represents a highly non-trivial test bench of our model which appear to perform very well.

\subsection*{Motifs} 

The motifs analysis has been carried on by calculating two different quantities. The first one has been defined as

\begin{equation}
s_m=\frac{N_m(\mathbf{M})-\langle N_m\rangle}{\langle N_m\rangle}
\end{equation}

\noindent and named \emph{similarity}: it quantifies the goodness of our prediction, measuring the difference between the observed and expected abundances. Beside similarity, we have also considered the traditional $z$-scores \cite{Caldarelli2007,Milo2002,mymotifs}, defined as the ratio of the difference between the observed and expected abundances and the corresponding standard deviation

\begin{equation}
z_m=\frac{N_m(\mathbf{M})-\langle N_m\rangle}{\sigma_m}
\end{equation}

\noindent with $\sigma_m=\sqrt{\langle N_m^2\rangle-\langle N_m\rangle^2}$ and $m$ indicating the particular motif considered. Even if $z$-scores have been recognized to be dependent on the network size \cite{Milo2004} (at least for monopartite networks), our dataset collects matrices with very similar volume ($R\in [0.56,0.61]$): thus, we can imagine this effect to be very small.  

Notice that similarity and $z$-scores provide complementary information: in particular, the latter measures the statistical significance of the agreement found by the former, accounting for the role of higher-order correlations not included in our constraints. Moreover, their ratio $s_m/z_m=\sigma_m/\langle N_m\rangle$ coincides with the motif-specific coefficient of variation, quantifying to what extent the average sums up the relevant information encoded into the corresponding ensemble distribution. Naturally, as for the 
observed abundances, both $s_m$ and $z_m$ can be defined for specific subsets of nodes as well.

\begin{figure*}[t!]
\begin{center}
\includegraphics[width=0.99\textwidth]{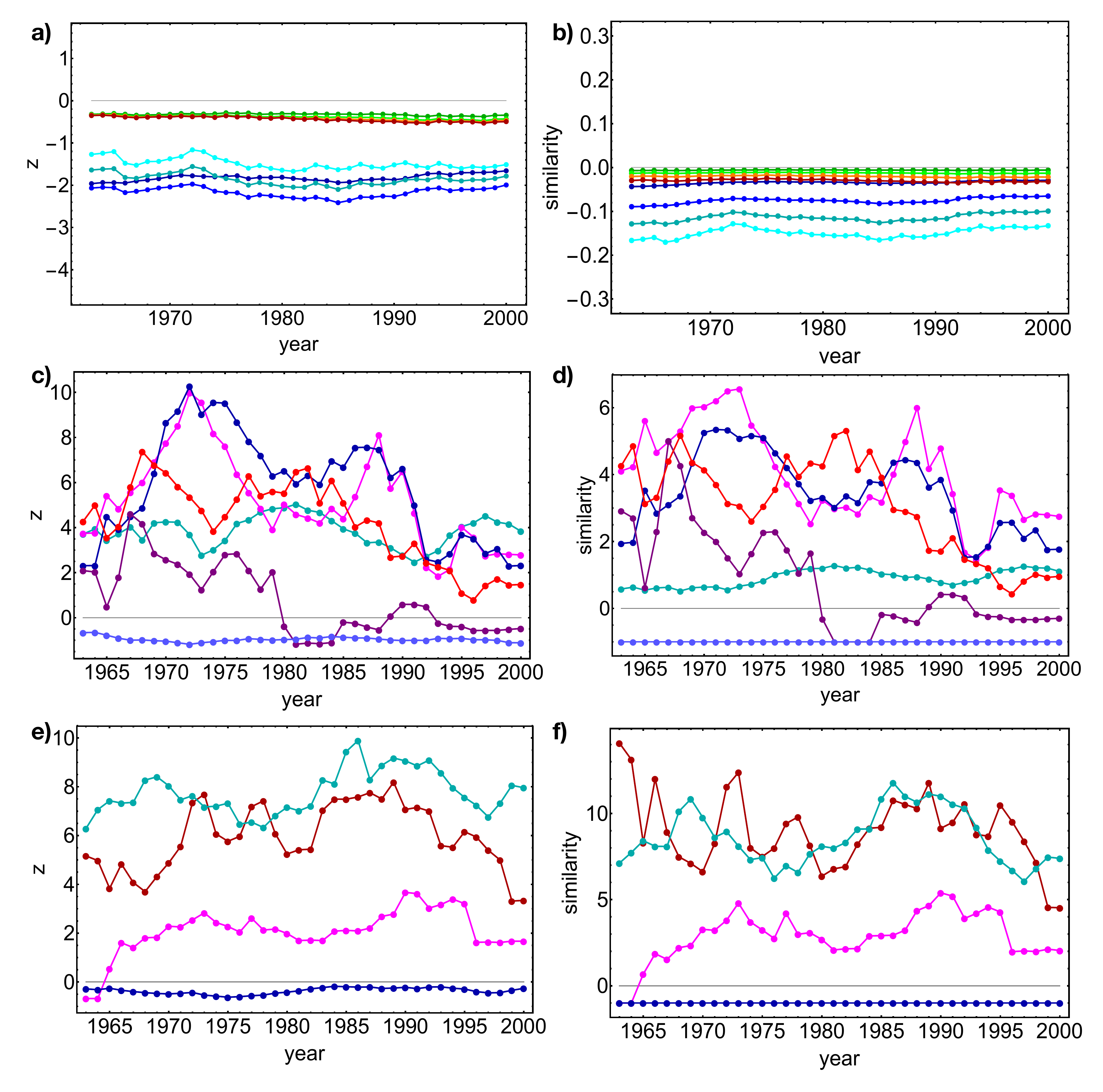}
\caption{Analysis of motifs. Top panels: $z$-scores (panel a) and similarity (panel b) evolution across our database years of $N_{V}$ ($\textcolor{Blue}{\bullet}$), $N_{V3}$ ($\textcolor{blue}{\bullet}$), $N_{V4}$ ($\textcolor{SeaGreen}{\bullet}$), $N_{V5}$ ($\textcolor{Cyan}{\bullet}$), $N_{\Lambda}$ ($\textcolor{Green}{\bullet}$), $N_{\Lambda3}$ ($\textcolor{green}{\bullet}$), $N_{\Lambda4}$ ($\textcolor{orange}{\bullet}$), $N_{\Lambda5}$ ($\textcolor{BrickRed}{\bullet}$). Middle panels: $z$-scores (panel c) and similarity (panel d) evolution of V$n$-motifs, restricted to subsets of countries - Asian Tigers ($\textcolor{Blue}{\bullet}$), Asian Tigers plus China ($\textcolor{Magenta}{\bullet}$), EU countries in G7 ($\textcolor{SeaGreen}{\bullet}$), BRICS ($\textcolor{Purple}{\bullet}$), eastern countries ($\textcolor{Red}{\bullet}$), four randomly chosen countries ($\textcolor{Cerulean}{\bullet}$). Bottom panels: $z$-scores (panel e) and similarity (panel f) evolution of $\Lambda n$-motifs, restricted to subsets of products - ``fruit and parts of plants'', ``aluminium and aluminium alloys'', ``road tractors'' ($\textcolor{Magenta}{\bullet}$), ``milk and cream'', ``butter'', ``cheese'' ($\textcolor{SeaGreen}{\bullet}$), four randomly chosen products ($\textcolor{Blue}{\bullet}$). Right column, panel f: similarity evolution across our database years of the same motifs. Our method correctly captures the countries tendency to expand their production ($\Lambda n$-motifs), even if the resemblance of the different baskets of products is overestimated (V$n$-motifs). Moreover, our method identifies statistically significative correlations among subsets of countries and products.}
\label{mot}
\end{center}
\end{figure*}

\begin{figure*}[t!]
\begin{center}
\includegraphics[width=0.99\textwidth]{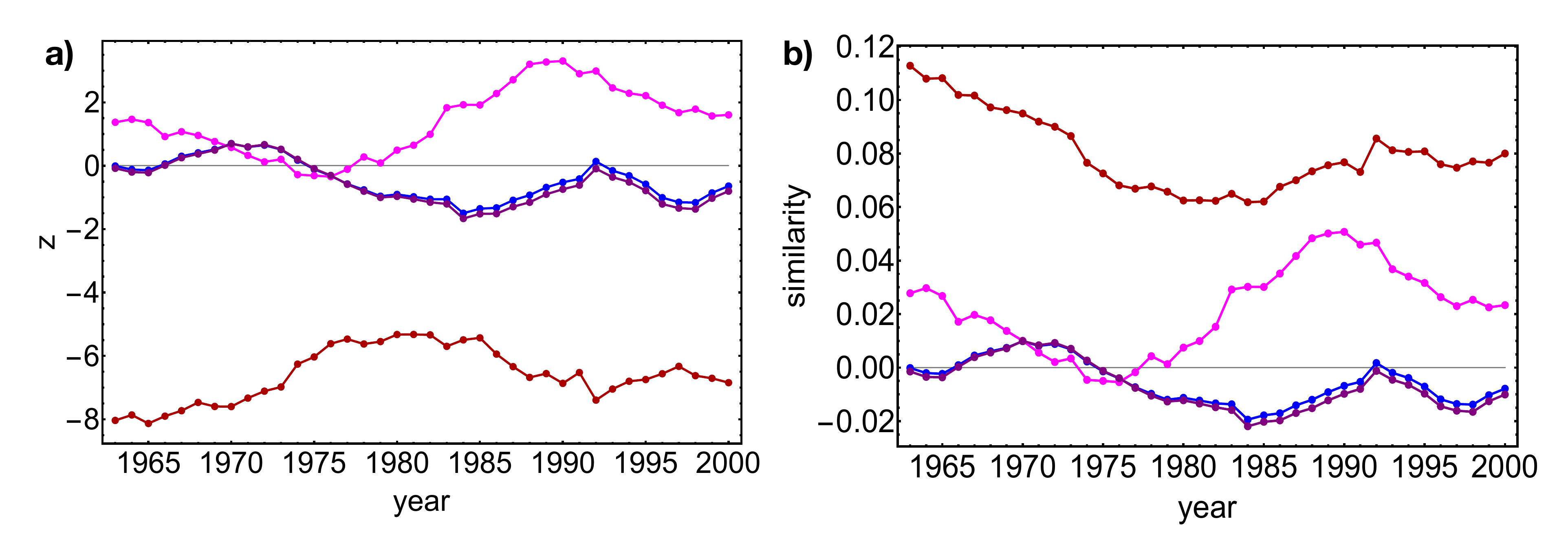}
\caption{Analysis of the assortativity coefficient and nestedness. $z$-scores (panel a) and similarity (panel b) evolution across our database years of $r$ ($\textcolor{BrickRed}{\bullet}$), NODF ($\textcolor{blue}{\bullet}$), nestedness along rows ($\textcolor{Magenta}{\bullet}$) and columns ($\textcolor{violet}{\bullet}$). While we are predicting a less disassortative network than observed, our method correctly reproduces the matrix nestedness.}
\label{rnest}
\end{center}
\end{figure*}

Fig. \ref{mot} shows the analysis of the V$n$ and $\Lambda n$ motifs. First, we have sampled the V-motifs and $\Lambda$-motifs abundance on the ensemble, in order to verify their distribution (see the Supplementary Information): both follow a gaussian very closely. Since all our motifs are sums of (neither independent nor identically distributed) random variables, this may be seen as a consequence of the generalized Central Limit Theorem. $z$-scores can be thus attributed the correct probabilistic meaning of (gaussian) standardized variables \cite{Milo2002,Milo2004} and choosing a threshold $z_0$ for $z$ allows the identification of significantly deviating patterns. In what follows we will choose $z_0=\pm 1.65$ as threshold values for the aggregated V$n$ and $\Lambda n$ families and $z_0=\pm2$ for the subsets-specific corresponding ones (see the Supplementary Information for a justification of such values). Naturally, if the observations were exactly reproduced by our null model, the $z$-scores would be zero.

The evolution of both similarity and $z$-scores across the years in our database point out that the $\Lambda n$ family is better reproduced than the V$n$ family (showing a similarity and a $z$-score closer to zero - see panels \ref{mot}a and \ref{mot}b). In particular, V$n$ $z$-scores lie outside the boundary of the significance region, showing values lower than $-1.65$. This indicates that for the binary, bipartite representation of the WTW, the degree sequence is far more effective in reproducing the products correlations than the correlations between countries. In other words, we correctly capture the countries tendency to expand their production, which seems to co-exist with a certain superposition of the countries baskets of products (see M-motifs in the Supplementary Information). However, the BiCM overestimates the resemblance of the different baskets: as $z$-scores indicate, world-countries tend to form less V-motifs than expected under our null model (further confirmed by the trend of X-motifs and W-motifs - see the Supplementary Information). Summing up, world countries show a clear tendency to diversify their production, at the same time avoiding to directly compete on the same products.

The comparison between similarity and $z$-score clarifies the role of average in characterizing the ensemble distribution of V$n$ and $\Lambda n$ families: the ratio $s_m/z_m \leq0.1$ justifies our interest in their ensemble average alone.

However, $z$-scores of V$n$ and $\Lambda n$ families result in almost flat trends which allow us to draw only general conclusions on the WTW as a whole. The reason lies in the ``aggregated'' character of such motifs, not distinguishing between different subsets of countries or products. To be more precise, let us consider the temporal evolution of our motifs on specific subsets of nodes (see panels \ref{mot}c and \ref{mot}d): in particular, the Asian Tigers (South Korea, Singapore, Taiwan, Hong Kong), the BRICS countries (Brazil, USSR/Russia, India, China, South Africa), the european countries belonging to G7 (France, Italy, Germany, United Kingdom) and a number of eastern-european countries (Hungary, Romania, Bulgaria, Poland, USSR/Russia) and let us calculate the temporal evolution of V4 and V5 motifs restricted to them. The european countries show a $z$-score almost constantly equal to 4, indicating a significant affinity which is maintained over time. An even stronger internal affinity is shown by the Asian Tigers to which China should be added (in fact, its addition to the group rises the $z$-score). On the other hand, BRICS countries show a very limited affinity \cite{Tacchella2012,Tacchella2013,Calda2012}: their trend becomes more and more consistent with the null model, to become negative in the recent years. The last two examples point out the limitations of the traditional economic classification (usually distinguishing China from Asian Tigers and gathering BRICS together), not capturing any actual economic likeness.

Eastern-european countries, on the other hand, show a strong correlation before 1989, gradually declining as this topical year approaches. Interestingly enough, after 1989 such correlation doesn't disappear, remaining statistically significant (and stabilizing around $z\simeq 2$): this seems to indicate a significant connection still persisting, having Russia replaced USSR as ``reference'' country. An additional test is provided by the random choice of four countries (Ghana, China, Mozambique, Austria): although close to zero, their trend is constantly negative. In fact, being Ghana and Mozambique low-diversification countries, they will be linked only to high-ubiquity products, common to all countries (see fig. \ref{knn2000}): thus, their basket will be far more limited than China's and Austria's, limiting in turn their possibility to compete. The constantly negative sign indicates, in this case, the impossibility to compete.

This kind of analysis can be repeated for $\Lambda n$ motifs as well, allowing us to gain a substantial insight into the products correlations. Panels \ref{mot}e and \ref{mot}f show some examples. While the food sector we have considered shows a constantly high value of $z$, indicating the common origin of the chosen dairy products, the pink trend signals a non-trivial positive correlation between the sectors represented by worked aluminium artifacts, tractors and fruit. A possible explanation may rest upon the consideration that tractors are constituted by parts in aluminium to be, in turn, used to transport the picked fruit. Consistently, the last group of products (cheese, rods and locomotives) is characterized by the value $z\simeq0$.

Notice that while for some groups of nodes the first moment encloses great part of the relevant information ($s_m/z_m \leq0.5$), for other groups higher-order moments could provide additional, useful information ($s_m/z_m \simeq1$), e.g. the distribution asymmetry. Interestingly, these circumstance are mostly encountered for countries and products, respectively.

\subsection*{Assortativity coefficient and nestedness}

As for the V$n$ and $\Lambda n$ motifs, the assortativity coefficient has a gaussian ensemble distribution (see the Supplementary Information). Both the observed value $r$ and its $z$-score signal that we are globally overestimating the network assortativity: more exactly, since our expected coefficient $\langle r\rangle$ is still negative, we are predicting a less disassortative network than observed (see fig. \ref{rnest}). This is a consequence of our randomization procedure, distributing links between nodes more homogeneously (recall that, consistently, our predicted $\{d^{nn}_p\}_{p=1}^P$ and $\{u^{nn}_c\}_{c=1}^C$ show less steeply decreasing trends than the observed ones - see fig. \ref{knn2000}). 

In order to better understand the concept of nestedness, let us explicitly draw a matrix from the BiCM-induced grandcanonical ensemble, ranking its rows and columns according to the $F_c$ and $Q_p$ measures \cite{Tacchella2012,Tacchella2013}. The result is shown in fig. \ref{rnest}. Notice that nestedness cannot be simply reduced to the concept of ``triangularity'' of a matrix. In fact, even if the drawn matrix shows a more curved boundary than the observed one, both the nestedness ensemble distribution (see the Supplementary Information) and its $z$-score (fig. \ref{rnest}) signal that our method reproduces it correctly.

\begin{figure}[t!]
\begin{center}
\includegraphics[width=0.49\textwidth]{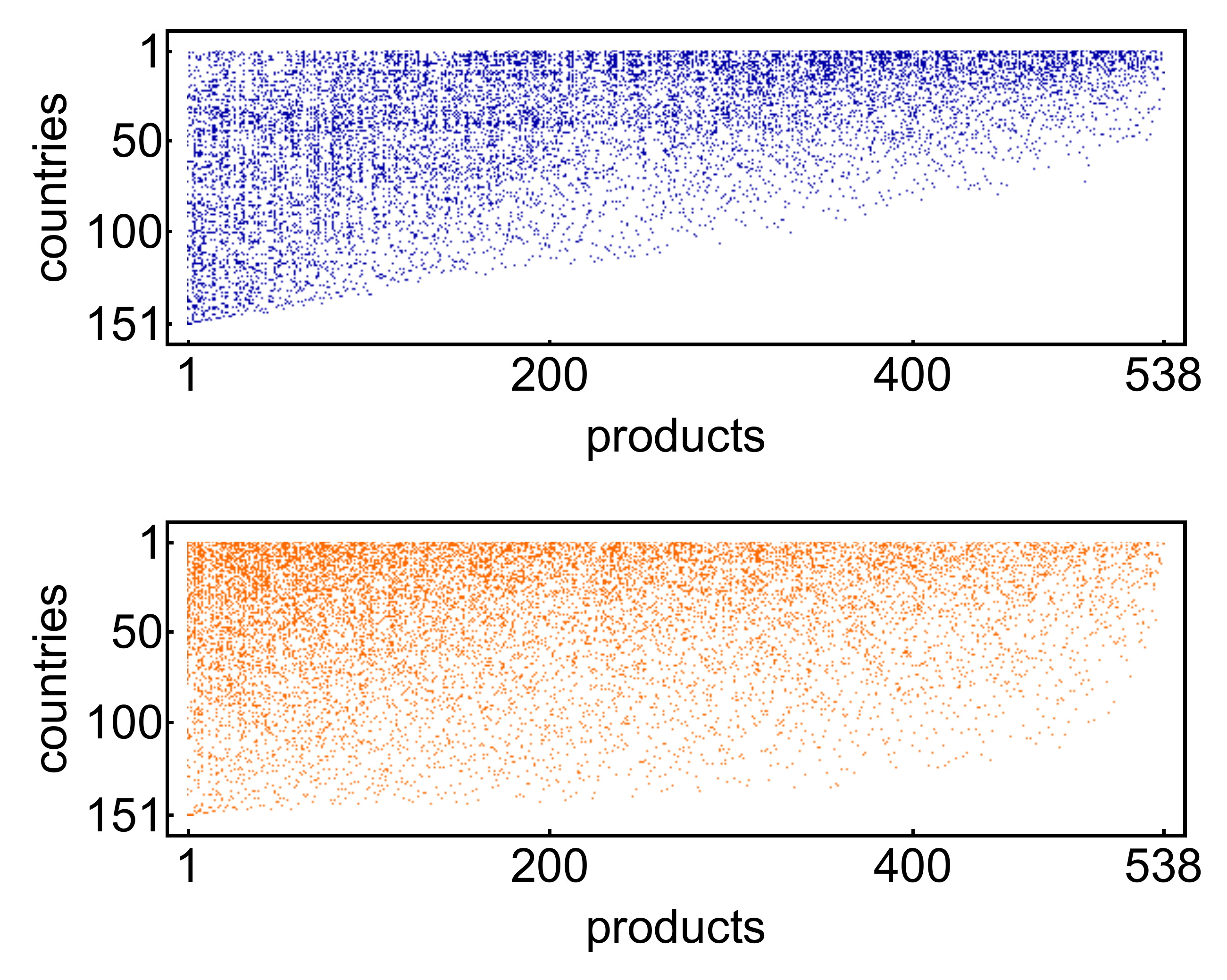}
\caption{Upper panel: the real World Trade Web matrix in the year 2000, with rows and columns in increasing order of fitness and complexity \cite{Tacchella2012,Tacchella2013}. Lower panel: matrix drawn from the BiCM-induced grandcanonical ensemble for the same year and ordered according to the same criterion.}
\label{rnest2}
\end{center}
\end{figure}

We have also measured the nestedness along rows and the nestedness along columns separately (according to the definitions in \cite{Almeida2008}). While the latter is reproduced and closely follows the trend of the global one, the former is, for a few years, significantly underestimated. This is non-trivially related to the way our null model redistributes V-motifs and $\Lambda$-motifs. However, as the bottom panel in fig. \ref{rnest2} suggests, a role seems to be played by the asymmetry of our bipartite matrix as well: in other words, the higher cardinality of the products layer seems to induce a preferential filling of the rows, making them more homogenenous and lowering their expected nestedness.

It should be also noted that the ensemble coefficient of variation for both $r$ and NODF show such a small value ($s_m/z_m\simeq 10^{-2}$ for both, across our temporal dataset) that the ensemble average can be considered as the only moment carrying relevant information.

\section*{Discussion}

In this paper we have both proposed a method to randomize binary, undirected, bipartite networks, by constraining essential network features as the total number of links and the nodes connectivity, and tested it on a real system as the World Trade Web. While, on the one hand, specifying the degree sequence allows highly non-trivial properties like countries fitness, products complexity and the matrix nestedness to become reproduced across our whole dataset, on the other quantities like assortativity and motifs still elude a satisfactorily explanation.

This is even more surprising, when considering the high level of accuracy achieved by the CM predictions in the analysis of the monopartite representation of the WTW. Our findings suggest that analysing different representations of the same network can indeed convey additional information, as proved by the agreement between the observed assortativity and the expected one (see fig. \ref{knn2000}), lower than in the corresponding monopartite WTW \cite{myPRE1}. In words, the correlations between countries induced by their productivity relations, clearly displayed by the bipartite representation of the WTW, are only partially explained by the degree sequence, calling for a higher amount of information to achieve the same level of accuracy obtained for the monopartite representation (and analogously for products). Otherwise stated, representing the same system via different network models (even belonging to the same class of binary, undirected configurations) may strongly affect the effectiveness of the corresponding piece of information (as the nodes connectivity) in reproducing the observed structure.

\begin{figure}[t!]
\begin{center}
\includegraphics[width=0.49\textwidth]{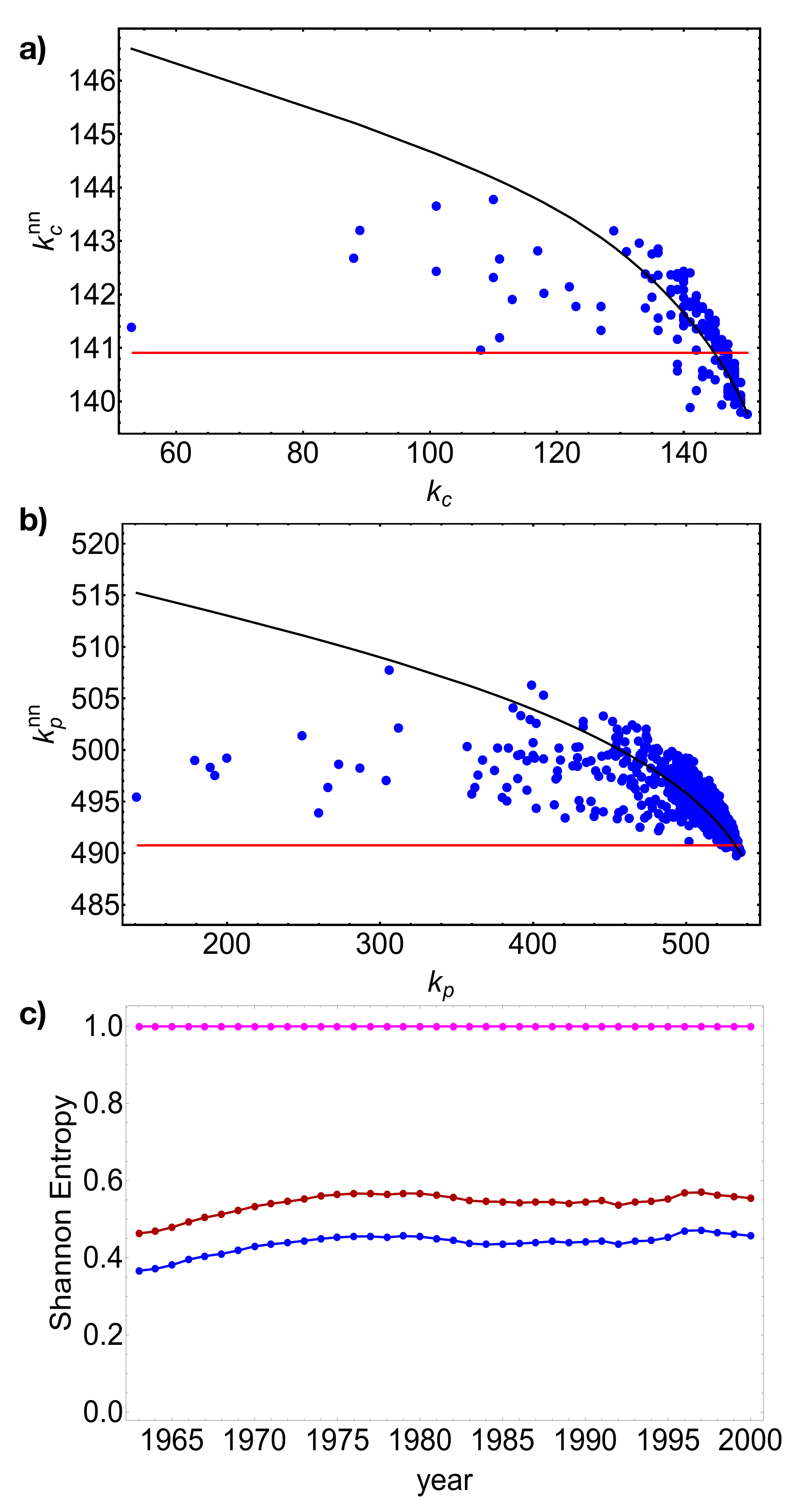}
\caption{Top panel: analysis of the degrees correlations on the projected WTW, in the year 2000, on the countries layer (blue: observed trend; black: prediction under the CM; red: prediction under the RG). Middle panel: analysis of the degrees correlations on the projected WTW, in the year 2000, on the products layer (blue: observed trend; black: prediction under the CM; red: prediction under the RG). Bottom panel: Shannon entropy of the uniform distribution ($\textcolor{Magenta}{\bullet}$), of the bipartite Random Graph model ($\textcolor{BrickRed}{\bullet}$) and of the bipartite Configuration Model ($\textcolor{blue}{\bullet}$) over the grandcanonical ensemble of binary, undirected, bipartite networks.}
\label{sha}
\end{center}
\end{figure}

Assortativity provides again the clearest example: as previously pointed out, the bipartite Configuration Model predicts trends quite similar to those expected under the bipartite Random Graph. To better quantify this difference, we have calculated the Shannon entropy (normalized to the total number of nodes pairs, i.e. the network volume) of the probability distributions induced by the BiRG and the BiCM:

\begin{equation}
S=\frac{-\sum_{c=1}^C\sum_{p=1}^P[p_{cp}\ln p_{cp}+(1-p_{cp})\ln (1-p_{cp})]}{C\cdot P}
\end{equation}

\noindent where $p_{cp}=\frac{x_cx_p}{1+x_cx_p}$ for the BiCM and $p_{cp}=\frac{x}{1+x}$ for the BiRG (see Supplementary Information). Results are shown in the bottom panel of fig. \ref{sha}. As evident from the trends, while specifying the total number of links strongly reduces the uncertainty (as signalled by the low value of the connectance, reducing the ensemble entropy to half its maximum value), further specifying the degree sequence produces a less relevant effect one could expect on the basis of the well known, monopartite results \cite{myPRE1}. Comparing the analyses of degree correlations for the bipartite and the projected WTW (on both countries and products layers - top and middle panels of fig. \ref{sha} for the year 2000), what emerges is quite impressive: while the CM prediction correctly overlaps to the observed trend, the RG predicts a flat trend completely missing the observed cloud of points (in line with the results already obtained for the monopartite representation \cite{myPRE1}). In terms of Shannon entropy, when passing from the RG to the CM the reduction of uncertainty on the observed, projected WTW amounts to $41\%$; for the bipartite WTW, this percentage reduces to only $16\%$ (see fig. \ref{sha}). This findings clearly indicate a future extension of our work: constraining those quantities having a significant impact on nodes correlations, as V-motifs, $\Lambda$-motifs or nestedness, in order to define a more effective null model.

However, as the analysis of motifs reveals, the BiCM provides the right benchmark to highlight meaningful correlations between countries and products, representing a purely topological alternative to the traditional economic classification, whose limitations have been already pointed out \cite{Tacchella2012,Tacchella2013,myintensive}. Remarkably, this kind of analysis can be repeated for different years, in order to monitor our system over time and detect significant temporal trends of the world economies co-evolution.

We stress that our approach is grandcanonical and possible extensions of the method move in the same direction. The paper in \cite{Tummi2011}, on the other hand, implements the microcanonical version of a mono-layer regular random graph: as for monopartite networks, comparing the performance of the two available approaches represents a challenging, future research direction.

Future work moves towards the direction of extending the present framework to directed, as well as weighted, networks models, to test the robustness of our findings also for configurations beyond the binary, undirected ones.

\section*{Supplementary Information}

\subsection*{The Random Graph model}

In the main text we have explicitly shown only the first and last passages of the calculations for the bipartite Configuration Model. The full passages are reported below:

\begin{eqnarray}
P(\mathbf{M}|\vec{\theta})&=&\frac{e^{-\sum_{c,p}(\alpha_c+\beta_p) m_{cp}(\mathbf{M})}}{\sum_{\mathbf{M}}e^{-\sum_{c,p}(\alpha_c+\beta_p) m_{cp}(\mathbf{M})}}\nonumber\\&=&\frac{\prod_{c,p}e^{-(\alpha_c+\beta_p) m_{cp}(\mathbf{M})}}{\sum_{\mathbf{M}}\prod_{c,p}e^{-(\alpha_c+\beta_p) m_{cp}(\mathbf{M})}}\nonumber\\&=&\frac{\prod_{c,p}(x_cy_p)^{m_{cp}(\mathbf{M})}}{\prod_{c,p}(1+x_cy_p)}\nonumber\\
&=&\prod_{c,p}p_{cp}^{m_{cp}}(1-p_{cp})^{1-m_{cp}}
\end{eqnarray}

\noindent (with $e^{-\alpha_c}= x_c$, $e^{-\beta_p}= y_p$ and $p_{cp}=\frac{x_cy_p}{1+x_cy_p}$). The calculations for the BiRG proceed along the same lines of those for the BiCM:

\begin{eqnarray}
P(\mathbf{M}|\theta)&=&\frac{e^{-\theta \sum_{c,p}m_{cp}(\mathbf{M})}}{\sum_{\mathbf{M}}e^{-\theta \sum_{c,p}m_{cp}(\mathbf{M})}}\nonumber\\&=&\frac{\prod_{c,p}e^{-\theta m_{cp}(\mathbf{M})}}{\sum_{\mathbf{M}}\prod_{c,p}e^{-\theta m_{cp}(\mathbf{M})}}
=\frac{x^{L(\mathbf{M})}}{\prod_{c,p}(1+x)}\nonumber\\
\end{eqnarray}

\noindent (with $e^{-\theta}= x$). Some more algebra leads to

\begin{eqnarray}
P(\mathbf{M}|x)&=&\left(\frac{x}{1+x}\right)^{L(\mathbf{M})}\left(\frac{1}{1+x}\right)^{C\cdot P-L(\mathbf{M})}\nonumber\\&=& p^{L(\mathbf{M})}(1-p)^{C\cdot P-L(\mathbf{M})}
\end{eqnarray}

\noindent with $x/(1+x)= p$. Maximizing the network log-likelihood function leads to the result $p=c(\mathbf{M})=L(\mathbf{M})/C\cdot P$.

\subsection*{Topological measures for binary, undirected, bipartite networks}

\paragraph*{Complexity and fitness.}

In order to infer the productive properties of the different countries from the biadjacency matrix \textbf{M}, in the context of Economic Complexity \cite{HH2009, Tacchella2012, Tacchella2013}, the fitness and complexity algorithm has been proposed in \cite{Tacchella2012}; roughly speaking, it is a generalization of the Google PageRank to bipartite networks. The algorithm assigns high fitness to the countries exporting the most exclusive (i.e. with higher complexity) products. In particular, the fitness $F_c$ for country $c$ and the complexity $Q_p$ for product $p$ are defined, at the $n$-th iteration of the algorithm, as
\begin{equation}\label{eq:FFQQ}
\left\{
\begin{array}{c}
\tilde{F}^{(n)}_c=\sum_{p=1}^P m_{cp} Q^{(n-1)}_p\\
\\
\tilde{Q}^{(n)}_p=\dfrac{1}{\sum_{c=1}^C m_{cp} \frac{1}{F^{(n-1)}_c}}
\end{array}
\right.
\rightarrow
\left\{
\begin{array}{c}
F^{(n)}_c=\dfrac{\tilde{F}^{(n)}_c}{\langle \tilde{F}^{(n)}_c\rangle}\\
\\
Q^{(n)}_p=\dfrac{\tilde{Q}^{(n)}_p}{\langle \tilde{Q}^{(n)}_p\rangle}
\end{array}
\right.,
\end{equation}
where the symbols $\langle\dots\rangle$ indicate the averages taken over the sets $\{\tilde{F}_c^{(n)}\}_{c=1}^C$ and $\{\tilde{Q}_p^{(n)}\}_{p=1}^P$. The initial conditions can be chosen to be $F_c^0=Q_p^0=1,\:\forall c,\:\forall p$. Further details on the convergence of the algorithm in (\ref{eq:FFQQ}) can be found in \cite{Pugliese2014}. The non-linear behaviour of fitness and complexity can be highlighted by respectively comparing the value of the diversification $d_c$ (ubiquity $u_p$) with the ranking obtained through the fitness $F_c$ (complexity $Q_p$) values, as shown in fig. 3 of the Main Text.

\paragraph*{Nestedness.}

Several different definitions of nestedness can be encountered in literature \cite{Almeida2008,Bastolla2009, Allesina2013, Munoz2013}. In the present article we use the definition called NODF (an acronym for \emph{Nestedness metric based on Overlap and Decreasing Fill}) and proposed in \cite{Almeida2008}. Let us define

\begin{equation}
S_{c\,c'}=\left\{
\begin{array}{c c}
d_c\neq d_{c'} & \dfrac{\sum_p\,m_{c\,p}\,m_{c'\,p}}{\text{min}\{d_c, d_{c'}\}}\\
&\\
\text{otherwise} & 0\\
\end{array}
\right.,
\end{equation} 
\begin{equation}
T_{p\,p'}=\left\{
\begin{array}{c c}
u_p\neq u_{p'} & \dfrac{\sum_c\,m_{c\,p}\,m_{c\,p'}}{\text{min}\{u_p, u_{p'}\}}\\
&\\
\text{otherwise} & 0\\
\end{array}
\right..
\end{equation} 

Notice that $S_{c\,c'}$ ($T_{p\,p'}$) are solely determined by those pairs of countries (products) for which the number of ones in rows $c$ and $c'$ (in columns $p$ and $p'$) are different. The measure of nestedness called NODF is then defined as
\begin{equation}\label{eq:nodfdef}
\text{NODF}=2\dfrac{\sum_{c<c'} S_{c\,c'}+\sum_{p<p'} T_{p\,p'}}{C(C-1)+P(P-1)}. 
\end{equation}
The definition (\ref{eq:nodfdef}) results from summing the contribution coming from rows and from columns, being normalized to the total number of couples of rows and columns. In order to isolate the single contributions coming from rows and columns, it is possible to defined the countries-specific and the products-specific NODF, respectively as

\begin{equation}
\text{NODF}_c=2\dfrac{\sum_{c<c'} S_{c\,c'}}{C(C-1)},\:\:\text{NODF}_p=2\dfrac{\sum_{p<p'} T_{p\,p'}}{P(P-1)}. 
\end{equation}

\subsection*{Expected topological measures for binary, undirected, bipartite networks}

\paragraph*{Assortativity.}

The expected value of the assortativity coefficients is easily calculable, after noticing that $\langle d_c\rangle=d_c$ and $\langle u_p\rangle=u_p$ by construction, that $m_{cp}^2=m_{cp}$, being $m_{cp}$ a binary variable, and resting upon the approximation $\langle\frac{n}{d}\rangle\simeq \frac{\langle n\rangle}{\langle d\rangle}$:

\begin{eqnarray}\label{exp:ANND}
\langle u^{nn}_c\rangle&=&\frac{\sum_{p=1}^Pp_{cp}(u_p-p_{cp}+1)}{d_c},\\
\langle d^{nn}_p\rangle&=&\frac{\sum_{c=1}^C p_{cp}(d_c-p_{cp}+1)}{u_p}.
\end{eqnarray}

Assortativity standard deviation can be calculated by applying the so-called \emph{delta method}, whose generic formula reads

\begin{equation}
\sigma_X\simeq\sqrt{\sum_{c=1}^C\sum_{p=1}^P\left(\frac{\partial X(\mathbf{M})}{\partial m_{cp}}\right)^2_{m_{cp}=p_{cp}}\sigma^2_{m_{cp}}},
\label{sigma}
\end{equation}

\noindent providing a method to calculate the standard deviation of any function of interest, $X(\mathbf{M})$, in terms of the independent random variables (in our case the entries $m_{cp}$ of the biadjacency matrix).

\begin{figure*}[t!]
\begin{center}
\includegraphics[width=0.99\textwidth]{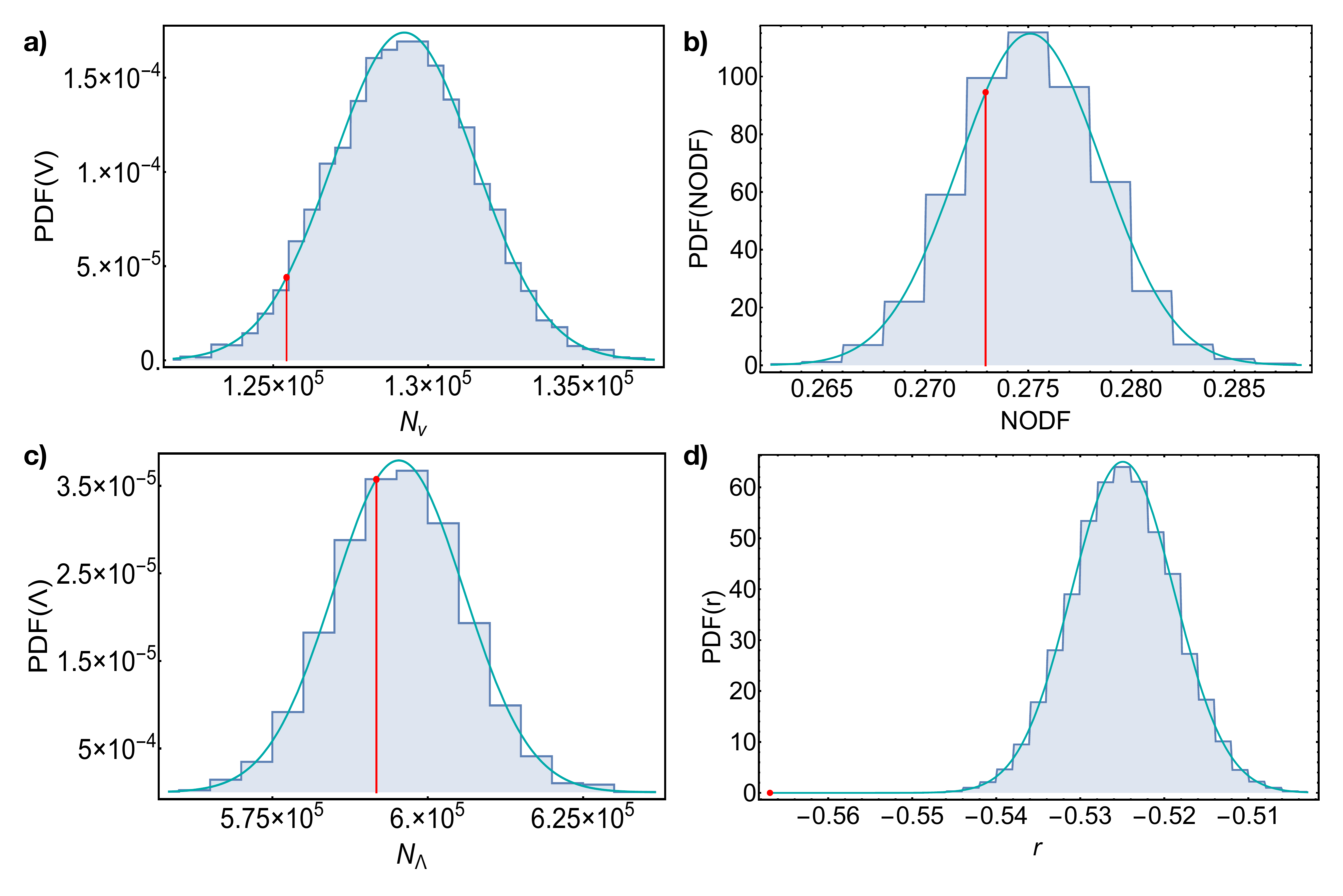}
\caption{Ensemble distribution of $N_{V}$ and $N_{\Lambda}$ abundance (panels a and c), assortativity coefficient $r$ and NODF (panels b and d) in the year 2000 (fits are obtained by superimposing normal distributions with the sample average and variance; red points represent the observed motifs abundance).}
\label{deg}
\end{center}
\end{figure*}

\paragraph*{Motifs.} 

In addition to the V$n$ and $\Lambda n$ family we can define more complex motifs, enlarging the number of nodes of the two layers to be considered. For example, X-\emph{motifs} can be defined, i.e. combinations of two V-motifs subtending the same pairs of countries and products (see fig. 2 in the Main Text): 

\begin{eqnarray}
N_{X}(\mathbf{M})&=&\sum_{c<c'}\sum_{p<p'}m_{cp}m_{cp'}m_{c'p}m_{c'p'}\nonumber\\&=&\sum_{c<c'}\binom{\mathcal{C}_{cc'}}{2}=\sum_{p<p'}\binom{\mathcal{P}_{pp'}}{2}
\end{eqnarray} 

\noindent (the notation $\sum_{c<c'}$ being equivalent to $\sum_{c=1}^C\sum_{c'=c+1}^C$ and similarly for products). As evident from the definition, X-motifs measure the co-occurrence of two countries as producers of the same pair of products and, viceversa, the co-occurrence of two products in the baskets of the same two countries. Thus, competitiveness on different segments of the market can now be measured, refining the information provided by V-motifs.

Allowing for an even higher number of nodes to interact, M-\emph{motifs} and W-\emph{motifs} can be defined (see fig. 2 in the Main Text) as 

\begin{eqnarray}
N_{M}(\mathbf{M})&=&\sum_{c<c'}\sum_{p<p'<p''}m_{cp}m_{cp'}m_{cp''}m_{c'p}m_{c'p'}m_{c'p''}\nonumber\\&=&\sum_{c<c'} \binom{\mathcal{C}_{cc'}}{3};\nonumber\\
N_{W}(\mathbf{M})&=&\sum_{p<p'}\sum_{c<c'<c''}m_{cp}m_{cp'}m_{c'p}m_{c'p'}m_{c''p}m_{c''p'}\nonumber\\&=&\sum_{p<p'}\binom{\mathcal{P}_{pp'}}{3}
\end{eqnarray}

\noindent respectively. As evident from the definition, countries competitiveness is now measured on a larger number of products.

Since all motifs are defined in terms of products of biadjacency matrix entries and the latter are treated as independent random variables by our null model, their expectation value can be computed exactly. Thus, we have

\begin{eqnarray}
\langle N_{V}\rangle&=&\sum_{c=1}^C\sum_{c'=c+1}^C\sum_{p=1}^P p_{cp}p_{c'p},\nonumber\\
\langle N_{\Lambda}\rangle&=&\sum_{p=1}^P\sum_{p'=p+1}^P\sum_{c=1}^C p_{cp}p_{cp'},\nonumber\\
\langle N_{X}\rangle&=&\sum_{c=1}^C\sum_{c'=c+1}^C\sum_{p=1}^P\sum_{p'=p+1}^Pp_{cp}p_{cp'}p_{c'p}p_{c'p'},\nonumber\\
\langle N_{M}\rangle&=&\sum_{c=1}^C\sum_{c'=c+1}^C\sum_{p<p'<p''}p_{cp}p_{cp'}p_{cp''}p_{c'p}p_{c'p'}p_{c'p''},\nonumber\\
\langle N_{W}\rangle&=&\sum_{p=1}^P\sum_{p'=p+1}^P\sum_{c<c'<c''}p_{cp}p_{cp'}p_{c'p}p_{c'p'}p_{c''p}p_{c''p'}.\nonumber\\
\end{eqnarray}

However, when computing the expected value of the generalizations of the V-motifs and $\Lambda$-motifs (the V$n$ and $\Lambda n$ families), higher-order powers of the nodes' degrees appear, since their definition reads $N_{Vn}=\sum_{p=1}^P \binom{u_p}{n}$ and $N_{\Lambda n}=\sum_{c=1}^C \binom{d_c}{n}$. In these cases, we can exploit the evidence that our degrees can be considered (with a good approximation) gaussian-distributed over the ensemble induced by the BiCM (for example, when considering $N_{V3}=\sum_{p=1}^Pu_p(u_p-1)(u_p-2)$, the well known result stating that odd central moments of a gaussian distribution are zero can be used to greatly simplify the calculations).

\begin{figure*}[t!]
\begin{center}
\includegraphics[width=0.92\textwidth]{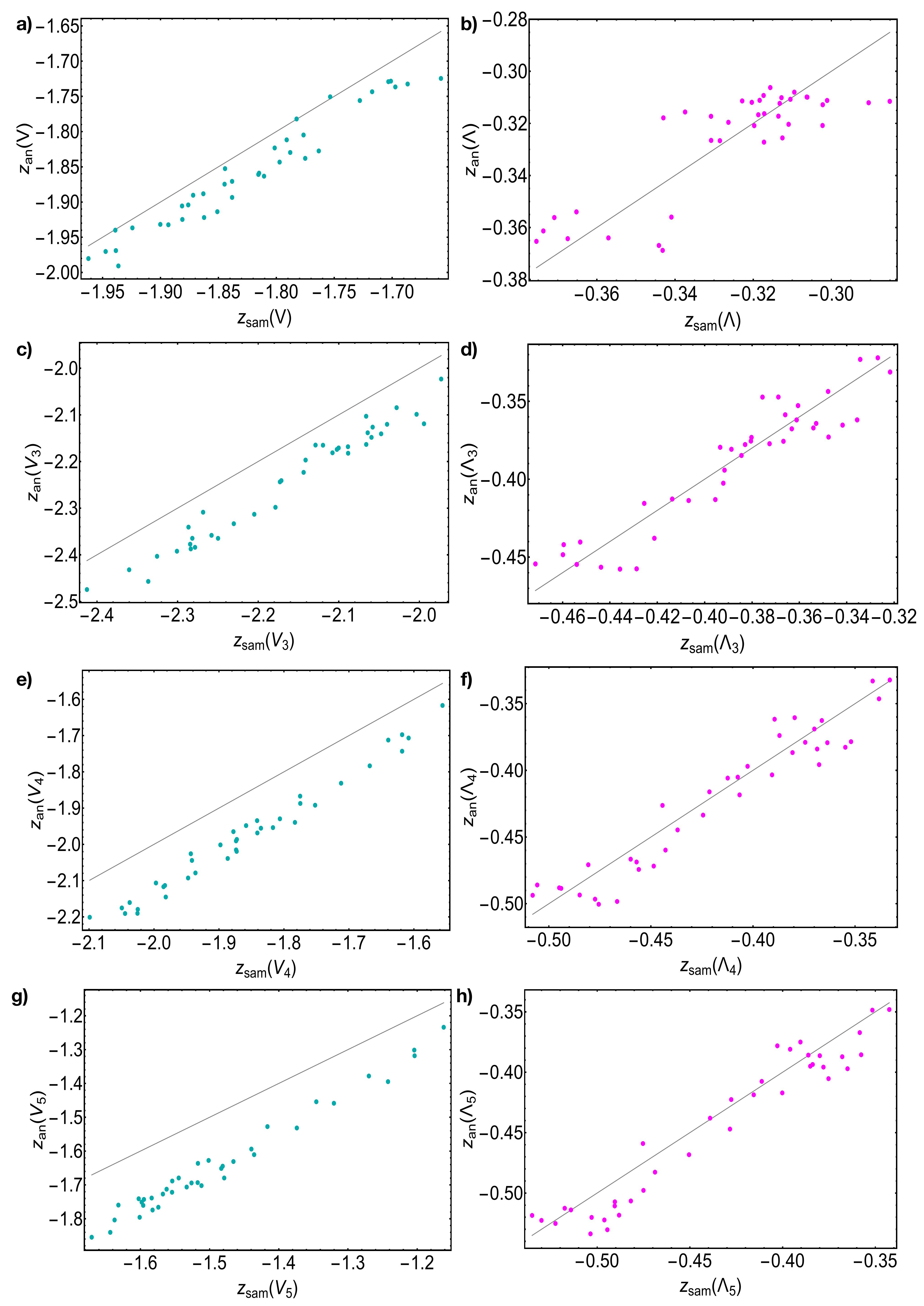}
\caption{Comparison between the analytical expressions of the $z$-scores (on the y-axis) and the corresponding values obtained by explicitly sampling the grandcanonical ensemble induced by the BiCM (on the x-axis), for $N_V$, $N_{V3}$, $N_{V4}$, $N_{V5}$ ($\textcolor{SeaGreen}{\bullet}$) and $N_\Lambda$, $N_{\Lambda3}$, $N_{\Lambda4}$, $N_{\Lambda5}$ ($\textcolor{Magenta}{\bullet}$).}
\label{zed}
\end{center}
\end{figure*}

The motifs standard deviation can be calculated by applying the delta method, with $X(\mathbf{M})=N_m(\mathbf{M})$. While valid in general, this formula assumes a particularly simple form for the V$n$ and $\Lambda n$ families of motifs. Expliciting it for a few cases will allow us to achieve a double goal: 1) providing a simple expression for the $z$-scores of the corresponding motifs and 2) showing a limitation of the traditional definition of $z$-scores. All the calculations will be carried on for the V$n$ family since they can be easily generalized to the $\Lambda n$ family. Let us start by noticing that V$n$ motifs are functions of the products degrees exclusively. Now, since in a bipartite network the nodes degrees within each layer are independent random variables, specifying eq. \ref{sigma} for the V$n$ family leads one to write

\begin{equation}
\sigma_{N_{Vn}}\simeq\sqrt{\sum_{p=1}^P\left(\frac{\partial N_{Vn}}{\partial u_p}\right)^2\sigma^2_{u_p}},
\label{motsigma}
\end{equation}

\noindent further simplifiable using the binomial result 

\begin{equation}
\frac{\partial \binom{u_p}{n}}{\partial u_p}=\binom{u_p}{n}(H_{u_p}-H_{u_p-n})
\end{equation}

\noindent with $H_i=\sum_{j=1}^i\frac{1}{j}$ being the \emph{i-th harmonic number}. Putting everything together, for $n=2$ we have

\begin{equation}
N_{V}=\sum_{p=1}^P \binom{u_p}{2}=\sum_{p=1}^P\frac{u_p(u_p-1)}{2};
\end{equation}

\noindent now, $\langle N_{V}\rangle=\sum_p\left[\langle u_p^2\rangle-u_p\right]/2$. This allows us to calculate the difference between $N_V$ and $\langle N_V\rangle$ simply as the total number of links variance

\begin{equation}
N_{V}-\langle N_{V}\rangle=-\frac{\sum_{p=1}^P\sigma^2_{u_p}}{2}=-\frac{\sigma^2_L}{2}
\end{equation}

\noindent where $\sigma^2_{u_p}=\langle u_p^2\rangle-\langle u_p\rangle^2=\sum_c\sigma^2_{m_{cp}}=\sum_cp_{cp}(1-p_{cp})$. In order to calculate the standard deviation, we use eq. (\ref{motsigma}) to find

\begin{equation}
\sigma^2_{N_V}\simeq\frac{\sum_{p=1}^P(2u_p-1)^2\sigma^2_{u_p}}{4}
\end{equation}

\noindent and finally obtain

\begin{equation}
z_V=\frac{N_V-\langle N_V\rangle}{\sigma_{N_V}}\simeq\frac{-\sigma^2_{L}}{\sqrt{\sum_{p=1}^P(2u_p-1)^2\sigma^2_{u_p}}}.
\end{equation}

The same procedure can be applied to all the motifs belonging to the V$n$ and $\Lambda n$ families. More explicitly, in the cases $N_{V3}=\sum_{p=1}^P \binom{u_p}{3}$ and $N_{V4}=\sum_{p=1}^P \binom{u_p}{4}$ the following results hold:

\begin{equation}
z_{V3}\simeq\frac{-\sum_p 3\sigma^2_{u_p}(u_p-1)}{\sqrt{\sum_{p=1}^P(3u_p^2-6u_p+2)^2\sigma^2_{u_p}}}
\end{equation}

\noindent and

\begin{equation}
z_{V4}\simeq\frac{-\sum_p [3\sigma^4_{u_p}+\sigma^2_{u_p}(6u_p^2-18u_p+11)]}{\sqrt{\sum_{p=1}^P(4u_p^3-18u_p^2+22u_p-6)^2\sigma^2_{u_p}}}.
\end{equation}

We have also tested the agreement between the analytical expressions of the aforementioned $z$-scores and the values obtained by explicitly sampling the grandcanonical ensemble induced by the CM. As fig. \ref{zed} shows, despite the presence of two approximations, our analytical estimates work quite satisfactorily.

\begin{figure}[t!]
\begin{center}
\includegraphics[width=0.49\textwidth]{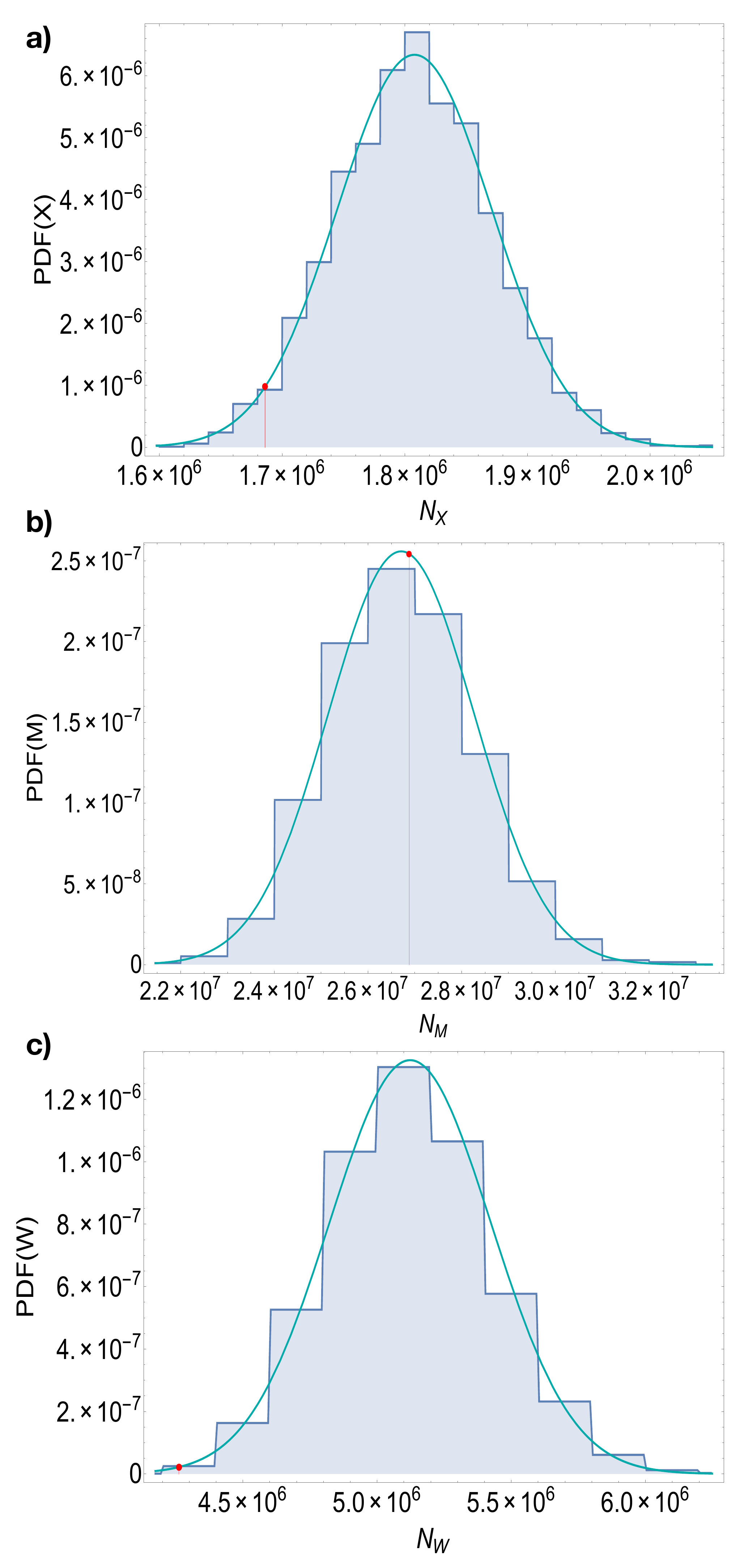}
\caption{Analysis of motifs. From top to bottom: ensemble distribution of $N_{X}$, $N_{M}$ and $N_{W}$ abundance in the year 2000 (fits are obtained by superimposing normal distributions with the sample average and variance; red points represent the observed motifs abundance).}
\label{mot4}
\end{center}
\end{figure}

\begin{figure}[t!]
\begin{center}
\includegraphics[width=0.49\textwidth]{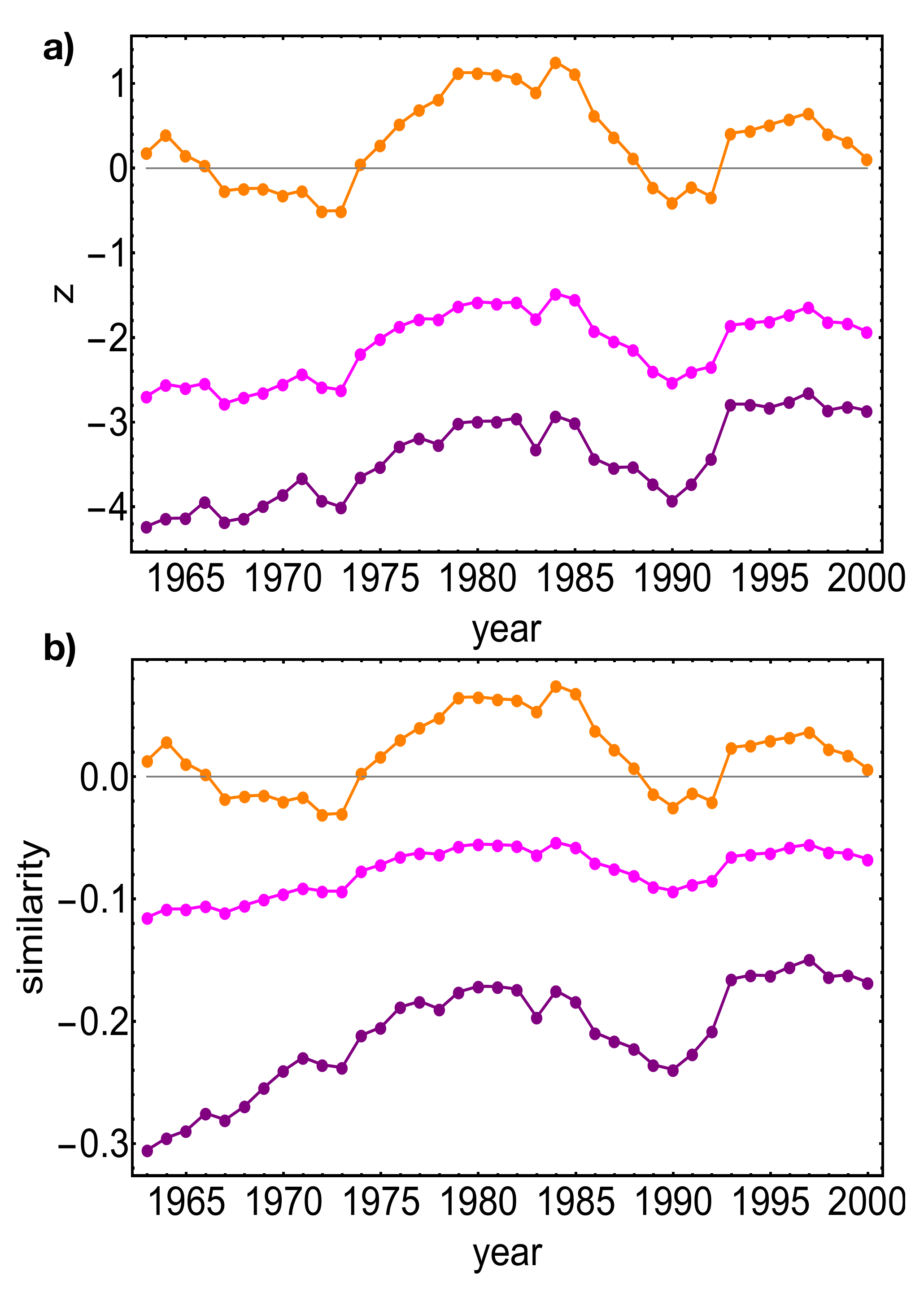}
\caption{Analysis of motifs. From top to bottom: $z$-scores and similarity evolution across our database years of $N_{M}$ ($\textcolor{orange}{\bullet}$), $N_{X}$ ($\textcolor{magenta}{\bullet}$), $N_{W}$ ($\textcolor{violet}{\bullet}$).}
\label{mot5}
\end{center}
\end{figure}

Figs. \ref{mot4} and \ref{mot5} show the analysis of the X, M and W-motifs. As for the other motifs previously considered, the three distributions follow a gaussian very closely, whose mean and variance have been calculated on the networks sample (5000 matrices). Again, this can be ascribed to the (generalized) Central Limit Theorem. As a general remark, the three, most complex motifs show higher fluctuations and are less accurately reproduced than the simpler ones (i.e. V$n$ and $\Lambda n$).

Beside having provided a simple expression for the $z$-scores of the V$n$ and $\Lambda n$ families of motifs, we have also shown that the latter may have a definite sign (negative, in our case): while quite surprising, in cases like this $z$-scores might still be used to test the agreement between observations and predictions but should be considered \emph{one-sided} statistical tests of significance. This implies that the values enclosing the probabilities of $68\%$, $95\%$ and $99\%$ no more coincide with $z=\pm 1$, $z=\pm 2$ and $z=\pm 3$, because no more computable on both tails of the reference gaussian distribution. The right $z$ values for one-sided tests are $\pm 1.65$, enclosing a probability of $95\%$, and $\pm 2.33$, enclosing a probability of $99\%$ \cite{Freund84}. The three more complex motifs (X-motifs, M-motifs and W-motifs) do not have a definite sign.

The reason for the sign definiteness lies in the explicit dependence of the quantities of interest from the chosen constraints, as shown below. Let us consider the Taylor expansion of a quantity of interest $f(x)$ around the expected value $\langle x\rangle$:

\begin{equation}
f(x)=f(\langle x\rangle)+\frac{\partial f}{\partial x}\Big\rvert_{\langle x\rangle}(x-\langle x\rangle)+\frac{\partial^2 f}{\partial x^2}\Big\rvert_{\langle x\rangle}\frac{(x-\langle x\rangle)^2}{2!}+\dots
\end{equation}

After calculating $\langle f\rangle$, the expression can be rewritten as

\begin{equation}
f(\langle x\rangle)-\langle f(x)\rangle=-\frac{\partial^2 f}{\partial x^2}\Big\rvert_{\langle x\rangle}\frac{\sigma^2_x}{2!}-\dots
\end{equation}

Whenever $x$ represents a given, chosen constraint whose expected value on the ensemble is, by definition, equal to the observed one, we obtain

\begin{equation}
f(x)-\langle f(x)\rangle=-\frac{\partial^2 f}{\partial x^2}\Big\rvert_{\langle x\rangle}\frac{\sigma^2_x}{2!}-\dots
\end{equation}

Now, if higher-order moments can be ignored or the function is quadratic in $x$, the right hand side of the above equation is proportional to the numerator of the $f$ function $z$-score, whose sign is negative. A simple example is provided by the function $x^2$, with $x=L$: we obtain $L^2-\langle L^2\rangle=\langle L\rangle^2-\langle L^2\rangle=-2\frac{\sigma^2_L}{2!}=-\sigma^2_L$, as it should. In the V-motifs case $f(\vec{x})=f(\{u_p\})=\sum_pu_p(u_p-1)/2$ and $N_V-\langle N_V\rangle=-\sum_p \sigma^2_{u_p}/2=-\sigma^2_L/2$ and analogously for the $\Lambda$, V3 and $\Lambda 3$ cases (remembering, for the latter, that odd central moments of a gaussian distribution are zero).

This finding has also an obvious interpretation in terms of grandcanonical and microcanonical ensembles. Given a certain set of constraints, the microcanonical approach prescribes them to be exactly satisfied, implying that no statistical fluctuations of the latter can be observed \cite{Tummi2011}. On the other hand, the grandcanonical approach cannot reduce such fluctuations to zero (as also clearly shown by fig. \ref{deg}) and the constraints variance will be positive: the dependence of a generic quantity of interest on it, according to the functional form shown before, is reflected in its sign definiteness.

\section*{Acknowledgments}
This work was supported by the EU project GROWTHCOM (611272) and the Italian PNR project CRISIS-Lab. The authors thank Giulio Cimini, Matthieu Cristelli and Andrea Tacchella for useful discussions.

\section*{Author Contributions}

FS and RDC analysed the data and prepared all figures. AG wrote the article. TS planned the research and wrote the article. All authors reviewed the manuscript.

\section*{Additional Information}

The authors declare no competing financial interests.

\end{document}